\documentclass[twocolumn,twocolappendix]{aastex702}

\usepackage[T1]{fontenc}
\usepackage{lmodern}
\usepackage{amsmath}
\newcommand{\spherex}{SPHEREx}
\newcommand{\tractorjax}{\textsf{TractorJAX}}
\newcommand{\thetractor}{\textsf{The Tractor}}
\newcommand{\eazy}{\textsf{eazy-py}}
\newcommand{\jax}{\textsf{JAX}}
\newcommand{\lsdr}{DESI Legacy Imaging Surveys DR10}
\newcommand{\ls}{LS}
\newcommand{\gaia}{\textit{Gaia}}
\newcommand{\euclid}{\textit{Euclid}}
\newcommand{\romantel}{\textit{Roman}}
\newcommand{\galsim}{\textsf{GalSim}}
\newcommand{\tjxspherex}{\textsf{tractorjax-spherex}}
\newcommand{\jaxversion}{0.9.0}
\newcommand{\helinewave}{1.083}
\newcommand{\psfcoreoffset}{0.05}

\newcommand{\efl}{\texttt{eigfloor}}
\newcommand{\eflp}{\texttt{eigfloor+prior}}
\newcommand{\lassoest}{\texttt{lasso}}
\newcommand{\linearest}{\texttt{linear}}
\newcommand{\mz}{\ensuremath{m_z}}
\newcommand{\zcut}{\ensuremath{\mz<21}}
\newcommand{\zbound}{\ensuremath{\mz=21}}
\newcommand{\crit}{\ensuremath{\mathcal{C}}}
\newcommand{\ncrit}{\ensuremath{N_{\mathcal{C}}}}
\newcommand{\fcrit}{\ensuremath{f_{\mathcal{C}}}}
\newcommand{\lassocut}{\ensuremath{\mz<20}}
\newcommand{\lassobound}{\ensuremath{\mz=20}}

\newcommand{\zspec}{\ensuremath{z_{\rm spec}}}
\newcommand{\zphot}{\ensuremath{z_{\rm phot}}}
\newcommand{\dz}{\ensuremath{\Delta z/(1+z)}}
\newcommand{\signmad}{\ensuremath{\sigma_{\rm NMAD}}}
\newcommand{\etaout}{\ensuremath{\eta_{0.15}}}
\newcommand{\um}{\ensuremath{\mu\mathrm{m}}}
\newcommand{\ujy}{\ensuremath{\mu\mathrm{Jy}}}

\newcommand{\nspecz}{4419}
\newcommand{\eigservicems}{11.9}
\newcommand{\hzcuteigms}{2.5}
\newcommand{\cpucoresec}{4.27}
\newcommand{\cpucoreseczcut}{0.70}
\newcommand{\hloadspread}{25\%}
\newcommand{\lampjobcores}{16}
\newcommand{\ntimingcut}{96}
\newcommand{\timingpix}{100}
\newcommand{\timingnsrc}{7520}
\newcommand{\timingnsrczcut}{1288}
\newcommand{\lfortyeigservicems}{20.3}
\newcommand{\lfortyzcuteigms}{3.0}
\newcommand{\cardgapfull}{1.7}
\newcommand{\cardgapzcut}{1.2}
\newcommand{\cpurenderfrac}{86\%}
\newcommand{\refcatsources}{5.8 billion}
\newcommand{\missionimages}{2.3 million}
\newcommand{\surveyyears}{two}
\newcommand{\activepix}{2040}
\newcommand{\cutoutsperimage}{416}
\newcommand{\missioncutouts}{$10^{9}$}
\newcommand{\missioncarddays}{130}
\newcommand{\missioncarddayszcut}{28}
\newcommand{\missioncoreyr}{130}
\newcommand{\missioncoreyrzcut}{21}
\newcommand{\dailycardhours}{four}
\newcommand{\nparity}{634}
\newcommand{\parityaltmed}{\ensuremath{10^{-4}}}
\newcommand{\npullmeas}{44{,}072}
\newcommand{\npullsrc}{3803}
\newcommand{\npullcut}{12}
\newcommand{\pullwhole}{0.999}
\newcommand{\pulltiled}{0.995}
\newcommand{\pullbelowthree}{97.5\%}
\newcommand{\npullsnrthree}{1080}
\newcommand{\pullsnrthree}{1.05}
\newcommand{\pullsnrthreepct}{5\%}
\newcommand{\sersicwingterm}{6.6\%}
\newcommand{\pullpoint}{0.996}
\newcommand{\sersiccorrresid}{1\%}
\newcommand{\npullprior}{7605}
\newcommand{\pullprior}{0.95}
\newcommand{\pullpriorexact}{0.94}
\newcommand{\pullpriornone}{0.91}
\newcommand{\pullpriorls}{0.96}
\newcommand{\pullpriormaxdiff}{0.06}
\newcommand{\pullpriormaxdiffci}{[+0.01, +0.10]}
\newcommand{\lspriornmad}{0.49}
\newcommand{\lspriormean}{+0.24}
\newcommand{\fptwoeigrel}{0.4\%}

\newcommand{\cutboxmin}{85}
\newcommand{\cutboxmax}{702}
\newcommand{\cutarcmin}{$8\farcm7$--$72\farcm0$}
\newcommand{\ncutmin}{308}
\newcommand{\ncutmax}{1002}
\newcommand{\ncuttotal}{4371}
\newcommand{\ndeepcut}{630}

\newcommand{\tilecore}{15}
\newcommand{\tilehalo}{3}
\newcommand{\eigfloorfrac}{\ensuremath{10^{-2}}}
\newcommand{\priorsigmafrac}{0.15}
\newcommand{\priorprecisioncap}{\ensuremath{10^{6}}}
\newcommand{\priorsigmamin}{\ensuremath{5\,\ujy}}
\newcommand{\priorslopemax}{2}
\newcommand{\ntemplates}{160}
\newcommand{\nboot}{20{,}000}

\newcommand{\tmplbumpfactor}{four}
\newcommand{\tmplfaintbumpfrac}{2\%}

\newcommand{\priorcatonly}{40}
\newcommand{\baselinecatonly}{12}
\newcommand{\pairedconfidence}{95\%}
\newcommand{\tmplwavelo}{1.3}
\newcommand{\tmplwavehi}{2.6}
\newcommand{\priorscanmin}{0.05}
\newcommand{\priorscanmid}{0.25}
\newcommand{\priorscanmax}{0.5}
\newcommand{\withhelddc}{\ensuremath{+0.54}}
\newcommand{\withhelddcci}{\ensuremath{[+0.05,+1.03]}}
\newcommand{\anchorpercentiles}{16th--84th}
\newcommand{\anchorfluxmax}{2000}
\newcommand{\anchorboot}{2000}
\newcommand{\lassozerofrac}{91--100}

\newcommand{\nrealiz}{24}
\newcommand{\depthwave}{2.40}
\newcommand{\truthchordratio}{1.07}
\newcommand{\truthbrightmag}{19}
\newcommand{\pullwaverange}{0.74--4.98}
\newcommand{\priorwrongsig}{0.5}
\newcommand{\popncommon}{337}
\newcommand{\popscatterratio}{1.34}
\newcommand{\popscatterpct}{34}
\newcommand{\popscatterci}{\ensuremath{[1.27,1.46]}}
\newcommand{\popscatterlinratio}{1.89}
\newcommand{\scatterrunglo}{0.96}
\newcommand{\scatterrunghi}{1.04}
\newcommand{\scatterfaintworse}{21}
\newcommand{\depthpercentiles}{16th--84th}
\newcommand{\plateaurange}{\text{20--22}}
\newcommand{\depthfaintbound}{22}
\newcommand{\huaizcut}{22.2}
\newcommand{\depthapradius}{10}
\newcommand{\cutmeasfrac}{60\%}
\newcommand{\fullmeasfrac}{20\%}
\newcommand{\cutfitn}{75}
\newcommand{\cutmeasn}{45}
\newcommand{\fullfitn}{227}
\newcommand{\fullmeasn}{46}
\newcommand{\deptheffratio}{5.5}
\newcommand{\modesliftcut}{2 of 636}
\newcommand{\modesliftfull}{298 of 3655}
\newcommand{\cuteigagree}{\ensuremath{4.8\times10^{-5}}}

\newcommand{\nlinearcommon}{3203}
\newcommand{\lineardnmad}{\ensuremath{+0.0001}}
\newcommand{\lineardnmadci}{\ensuremath{[-0.0001,+0.0003]}}
\newcommand{\lineardiscord}{seven}
\newcommand{\ntier}{2253}
\newcommand{\ntierfull}{1357}
\newcommand{\critlosspct}{40}
\newcommand{\tierdeliv}{0.0080}
\newcommand{\tierfulldeliv}{0.0078}
\newcommand{\cosmosshare}{55.7\%}
\newcommand{\ncomaext}{2899}
\newcommand{\comashare}{37\%}
\newcommand{\nlowzext}{965}
\newcommand{\ncomalowz}{773}
\newcommand{\linearmcnemar}{0.0156}
\newcommand{\ntierprior}{2182}
\newcommand{\sciencemedsnr}{2.27}
\newcommand{\fullmedsnr}{1.06}
\newcommand{\medsnrdrop}{2.1}
\newcommand{\simsnrdrop}{1.6}
\newcommand{\nsimretcommon}{635}
\newcommand{\simretthreshold}{3}
\newcommand{\simretreported}{1.55}
\newcommand{\simretactual}{1.42}
\newcommand{\bahktierpub}{0.015}
\newcommand{\bahktierhere}{0.0080}
\newcommand{\lassodnmad}{\ensuremath{0.0000}}
\newcommand{\lassodnmadci}{\ensuremath{[-0.0004,+0.0003]}}
\newcommand{\lassodcat}{\ensuremath{+0.40}}
\newcommand{\lassodcatci}{\ensuremath{[+0.10,+0.69]}}
\newcommand{\lassodiscorda}{16}
\newcommand{\lassodiscordb}{4}
\newcommand{\lassofullncommon}{3036}
\newcommand{\lassofulldnmad}{\ensuremath{-0.0002}}
\newcommand{\lassofulldnmadci}{\ensuremath{[-0.0006,+0.0002]}}
\newcommand{\lassofulldcat}{\ensuremath{+0.23}}
\newcommand{\lassofulldcatci}{\ensuremath{[-0.10,+0.56]}}
\newcommand{\lassofulldiscorda}{17}
\newcommand{\lassofulldiscordb}{10}
\newcommand{\lassofulldretention}{\ensuremath{+1.77}}
\newcommand{\lassodretention}{\ensuremath{+1.36}}
\newcommand{\lassogainn}{575}
\newcommand{\lassogainnmad}{\ensuremath{0.25}}
\newcommand{\lassogaincat}{56\%}
\newcommand{\lassobothcatlasso}{4.5\%}
\newcommand{\lassobothcateig}{3.0\%}
\newcommand{\lassobothn}{3171}
\newcommand{\lassopenn}{135}
\newcommand{\lassopennmadeig}{0.054}
\newcommand{\lassopennmadlasso}{0.26}
\newcommand{\lassopencateig}{22\%}
\newcommand{\lassopencatlasso}{51\%}
\newcommand{\priordnmad}{\ensuremath{+0.0031}}

\newcommand{\atefsym}{\ensuremath{A_{\rm TEF}}}
\newcommand{\atefshipped}{\ensuremath{0.70}}
\newcommand{\atefopt}{\ensuremath{0}}
\newcommand{\atefoptprior}{\ensuremath{0}}
\newcommand{\npriorpairnine}{3034}
\newcommand{\npriorfieldsnonsparse}{six}
\newcommand{\priordnmadincrease}{0.0015}
\newcommand{\priordnmadadopted}{\ensuremath{+\priordnmadincrease}}
\newcommand{\priordnmadadoptedci}{\ensuremath{[+0.00098,+0.00215]}}
\newcommand{\priordcatadopted}{\ensuremath{+0.92}}
\newcommand{\priordcatadoptedci}{\ensuremath{[+0.46,+1.38]}}
\newcommand{\priorsharedfrac}{52\%}

\newcommand{\ncosmosext}{4337}

\newcommand{\snpresel}{\ensuremath{\mathrm{S/N}>3}}
\newcommand{\nexternal}{7781}

\newcommand{\bahkpsfradius}{\ensuremath{1\arcsec}}
\newcommand{\bahkpsfbias}{5--10\%}
\newcommand{\bahkpsfbiasmax}{20\%}

\newcommand{\allpsfextloss}{32--40\%}
\newcommand{\allpsfextradius}{\ensuremath{3\arcsec}}

\newcommand{\snweightfunction}{\operatorname{clip}\!\left[1.25+0.25\frac{s-10}{20},\,1.25,\,1.50\right]}
\newcommand{\bahkpullwidth}{1.4}
\newcommand{\atefgridmax}{0.70}
\newcommand{\atefplateaulevel}{95\%}
\newcommand{\nsparsemin}{20}

\newcommand{\snonecross}{19.5}
\newcommand{\snthreecross}{21.2}
\newcommand{\snthreerange}{20.8--22.4}

\newcommand{\pedblind}{\ensuremath{+13.4}}
\newcommand{\pedprior}{\ensuremath{+9.2}}
\newcommand{\pedwithheld}{\ensuremath{+9.7}}
\newcommand{\wtwonfields}{nine}
\newcommand{\wtwonresolved}{seven}
\newcommand{\gwoneblind}{0.995}
\newcommand{\gwoneprior}{0.962}
\newcommand{\gwtwoblind}{0.982}
\newcommand{\gwtwoprior}{0.914}

\newcommand{\pairedcalratio}{1.31}
\newcommand{\pairedcalratiolo}{1.23}
\newcommand{\pairedcalratiohi}{1.41}
\newcommand{\npairedcal}{1992}
\newcommand{\pairedbinlo}{1.2}
\newcommand{\pairedbinhi}{1.6}
\newcommand{\pairedcritratio}{1.17}
\newcommand{\pairedcritratiolo}{1.09}
\newcommand{\pairedcritratiohi}{1.26}
\newcommand{\npairedcrit}{1354}
\newcommand{\pairedcritpct}{17}
\newcommand{\medmzcrit}{18.1}
\newcommand{\medmzcritfull}{17.6}
\newcommand{\lostmedmz}{18.5}
\newcommand{\lostnmadcut}{0.0110}
\newcommand{\ncritlost}{896}
\newcommand{\critlostfrac}{79\%}
\newcommand{\critlostmaglo}{18}
\newcommand{\pairedbinmax}{19.5}

\newcommand{\amratiolo}{1.6}
\newcommand{\amratiohi}{1.8}
\newcommand{\amsigmalo}{0.006}
\newcommand{\amsigmahi}{0.011}

\newcommand{\labdsigmax}{0.0003}
\newcommand{\labdznmadlo}{0.0006}
\newcommand{\labdznmadhi}{0.0007}
\newcommand{\labpriordnmad}{\ensuremath{+0.0016}}
\newcommand{\labpriordnmadci}{\ensuremath{[+0.00101,+0.00226]}}
\newcommand{\labpairedcritratio}{1.16}
\newcommand{\labpairedcritratiolo}{1.07}
\newcommand{\labpairedcritratiohi}{1.26}
\newcommand{\labfluxmed}{0.06\%}
\newcommand{\labrtol}{10\%}

\newcommand{\warpfoldbiaslo}{\ensuremath{-0.00072}}
\newcommand{\warpfoldbiashi}{\ensuremath{+0.00052}}

\shorttitle{\spherex{} Forced Photometry with \textup{\tractorjax{}}}
\shortauthors{Bahk et al.}

\begin{document}

\title{\tractorjax{}: GPU-Accelerated \spherex{} Forced Photometry\\
and the Trade-offs of Regularization and External Priors}

\author[0009-0002-9878-1126]{Hyeonguk Bahk}
\affiliation{Astronomy Program, Department of Physics and Astronomy, Seoul
National University, 1 Gwanak-ro, Gwanak-gu, Seoul 08826, Republic of Korea}
\email[show]{bahkhyeonguk@gmail.com}

\author[0000-0003-3428-7612]{Ho Seong Hwang}
\affiliation{Astronomy Program, Department of Physics and Astronomy, Seoul
National University, 1 Gwanak-ro, Gwanak-gu, Seoul 08826, Republic of Korea}
\affiliation{SNU Astronomy Research Center, Seoul National University,
1 Gwanak-ro, Gwanak-gu, Seoul 08826, Republic of Korea}
\affiliation{Institute for Data Innovation in Science, Seoul National
University, Seoul 08826, Republic of Korea}
\email[show]{hhwang@astro.snu.ac.kr}

\begin{abstract}
We present \tractorjax{}, an open-source GPU reimplementation of the
catalog-forced mode of \thetractor{}. We use it to test whether
regularization or external flux priors enable reliable fitting of
reference catalogs that extend below the depth directly constrained by
undersampled \spherex{} images. For full-depth regularized fits,
\tractorjax{} fits $\timingpix\times\timingpix$-pixel cutouts at
\eigservicems{}\,ms each on a single NVIDIA H100 GPU.
In simulated images of the Abell 2537 galaxy-cluster field, extending the
fitted catalog from \zcut{} to the full \lsdr{} catalog increases the
scatter in fractional flux residuals by \popscatterpct{}\%,
even with regularization. We also test the methods using \spherex{}
observations in nine fields, assessing the number of reliably measured sources and
photometric-redshift accuracy. Requiring $\mathrm{S/N}>5$ in more than
half of the 102 channels, we find that the regularized full-depth fit
yields \critlosspct{}\% fewer sources than the truncated fit. For sources
meeting this requirement in both fits, the normalized median absolute
deviation of the photometric redshifts, \signmad{}, is \pairedcritpct{}\%
larger in the full-depth fit. External flux priors recover nearly as many
sources as the \zcut{} fit, but increase \signmad{} by
\priordnmadincrease{} for the common calibrated sample. Neither
regularization nor the external flux priors tested here simultaneously
preserve the sample size and photometric-redshift accuracy achieved by
the truncated fit. Our results suggest that the optimal fitted-catalog
depth cannot be determined from image-level performance alone, but should
be selected through end-to-end tests tailored to the intended
science measurement.
\end{abstract}


\section{Introduction}\label{sec:intro}

\spherex{} measures a low-resolution spectrum of every source on the sky
using forced photometry against a deep external catalog. The mission images
the full sky repeatedly in 102 spectral channels between 0.75 and 5.0\,\um{}
with $6\farcs15$ pixels \citep{dore2014spherex,bock2025spherex}, which
are wider than the point spread function (PSF) full width at half maximum (FWHM) in every band
\citep{crill2025skysim,korngut2026instrument}; the survey is undersampled by
design. Because the pixel, rather than the PSF, sets the resolution
element, neighboring sources blend within individual pixels. Detection and deblending
from the images alone are therefore severely limited. The science
pipeline instead fixes source positions and profiles from a reference
catalog assembled from deeper ancillary surveys and fits only the
per-channel fluxes \citep{akeson2025pipeline,crill2025skysim}. This
approach enables photometric redshifts for hundreds of millions of
galaxies \citep{stickley2016spherex,feder2024spherex}. The accuracy of these redshifts
depends on how well
a forced fit behaves when the catalog is far deeper than the images.

Forced photometry against a deeper, higher-resolution catalog is a mature
framework. \thetractor{} \citep{lang2016tractor} models every cataloged
source simultaneously as a PSF-convolved profile and solves for the fluxes
that best explain the pixels. Catalogs built with it include the unWISE forced photometry
\citep{lang2014unwise,lang2016wise}, the DESI Legacy Imaging Surveys catalogs
\citep{dey2019legacysurveys}, and COSMOS2020 \citep{weaver2023farmer}. The
same idea constrains fits to confusion-limited infrared maps
\citep{laidler2007tfit,merlin2015tphot,labbe2015irac} and, in the case
closest to \spherex{}, recovers light curves from undersampled Kepler and
TESS pixels against a \gaia{} prior list \citep{hedges2021lfd,han2023tglc}.
The key feature \spherex{} needs is shared by these approaches: jointly
fitting neighboring sources rather than correcting for their light
afterward, and propagating the resulting uncertainty into the reported
errors.

At \spherex{} scale, this framework meets two challenges. The first is
computational; the reference catalog holds about \refcatsources{} sources,
each observed in 102 channels over repeated all-sky surveys that
together yield about \missionimages{} spectral images
\citep{crill2025skysim}. Because a full-depth forced fit of a single
spectral image takes tens of minutes on one CPU core, repeating it over the
survey is costly whenever the calibration, the reference catalog, or the
science goal changes. For example, studies of the galaxy clusters
known across the whole sky \citep{bahk2024upclustersz} may require
photometry tailored to their crowded fields.

Fortunately, this computation is well
suited to GPUs. With source positions and profiles fixed, each fit is
linear in the fluxes and reduces to dense linear algebra that can be
batched over many images. Existing GPU photometry codes
\citep{melchior2018scarlet,melchior2026scarlet2,stone2023astrophot,johnson2024forcepho}
achieve large speedups but fit nonlinear source models one scene at
a time. In contrast, the codes closest to the \spherex{} problem fit
confusion-limited maps against a list of known source positions but run on
CPUs \citep[e.g., XID+;][]{hurley2017xid}.
\spherex{} therefore needs a tool that is narrower in scope and larger in scale:
the linear catalog-forced mode of \thetractor{}, on which the \spherex{}
pipeline already rests, implemented on GPUs and batched over millions of
images.

The second challenge is statistical; an undersampled \spherex{} image
cannot independently constrain the fluxes of all sources in a catalog as
deep as the \lsdr{} (hereafter \ls{}). Bright sources can be
reliably deblended, but the information available for individual sources
becomes insufficient near the detection limit of the survey, beyond which most of the
catalog lies. The
joint fit is then severely ill-conditioned; its solution wanders through a
near-degenerate subspace in which flux fragments across unresolved neighbors, even though the residuals stay statistically flat. This problem differs from classical
confusion, which describes the noise contributed by sources omitted from
the model \citep{condon1974confusion,hogg2001confusion}; here, the bias
instead arises from sources that are included in the fit even though their
fluxes cannot be separately constrained \citep{melchior2021blending}.

Two simulation studies have examined this problem for \spherex{}.
\citet{huai2026confusion}
simulated the resulting trade-off between unmodeled confusion and
blended-fit errors as a function of reference-catalog depth and adopted a
magnitude cut. \citet{bahk2026clusters} used simulations of galaxy clusters,
which are the densest environments that \spherex{} will observe, and found that source
blending is the primary cause of catastrophic photometric-redshift outliers. They identified
pre-selecting the fitted catalog and regularizing the fit as two possible
ways to control this degeneracy, but did not determine which approach
performs best.

The simplest control is to truncate the fitted catalog at a magnitude limit,
keeping only sources bright enough to be individually constrained. This
approach is already used in confusion-limited infrared surveys, where the
prior list is either limited to sources that the map can constrain
\citep{roseboom2010xid} or selected using predicted spectral energy
distributions (SEDs) \citep{liu2018superdeblended,jin2018superdeblended}.
Both previous \spherex{} studies adopted a similar strategy;
\citet{huai2026confusion} applied the magnitude cut described above,
whereas \citet{bahk2026clusters} used a pre-selection and suggested a
broadband selection criterion based on the reference catalog. Although truncation
stabilizes the fit, the removed population does not leave
the data; its light remains in the pixels, unrepresented by the model, and biases the
surviving targets \citep{eddington1913bias,hoggturner1998faint}.

The alternative, also developed for confusion-limited maps, is to keep the
full catalog and control the degeneracy inside the fit. One option is a
ridge-type penalty, which can stabilize an ill-conditioned fit without
removing sources
\citep{tikhonov1963regularization,hoerl1970ridge}. Another is the use of
explicit per-source flux priors, including priors predicted from external
photometry, as in fits to \textit{Herschel} maps
\citep{leiton2015goodsherschel,hurley2017xid,pearson2017deblending};
a third, sparse $\ell_1$ selection, has been applied in the same setting
\citep{wang2014desphot,tibshirani1996lasso}. Whether that alternative works
at \spherex{} density remains untested, in simulation or in observational data.

This paper addresses both challenges. We present \tractorjax{}, a GPU
reimplementation of the catalog-forced mode of \thetractor{} in \jax{}
\citep{jax2018github}, and test its reported uncertainties against
simulation truth. In \tractorjax{}, regularization, external flux priors, and
sparse selection are configurations of one solver. We use \tractorjax{}
to quantify the cost of fitting beyond the depth supported by the data.
In truth-controlled simulations, we determine how many source fluxes the
pixels can constrain and assess the consequences of fitting additional
sources. Using observational data from nine fields, we compare two
approaches for extending the fit beyond this limit (regularization alone
and external flux priors) through their impact on photometric redshifts
in a fixed comparison sample. Photometric redshifts are the end-to-end
test of the 102-channel spectrophotometry; physical parameters from SED
fitting are outside the scope of this work.

Section~\ref{sec:data} describes the observational data and the comparison
sample. Section~\ref{sec:method} describes the engine, its configurations,
the photometric-redshift chain, and the simulations. We present the
results in Section~\ref{sec:results}, discuss them in
Section~\ref{sec:discussion}, and summarize the paper in
Section~\ref{sec:summary}. We quote magnitudes in the AB system and fluxes
in mJy throughout unless noted.

\section{Data and Samples}\label{sec:data}

\subsection{SPHEREx Images and the Reference Catalog}\label{sec:images}

Our \spherex{} observational sample consists of the COSMOS field and
eight galaxy-cluster fields from \citet{bahk2026clusters}: A1361, A2055, A2187, A2537, Coma,
SpARCS~J1613$+$5649, SPT-CL~J0546$-$5345, and SPT-CL~J2145$-$5644. We use
Quick Release 2 (QR2) Level~2 products \citep{akeson2025pipeline,irsa2025qr2}, queried
and downloaded from the NASA/IPAC Infrared Science Archive (IRSA)
collections \texttt{spherex\_qr2} and \texttt{spherex\_qr2\_deep} with
\texttt{astroquery}
\citep{ginsburg2019astroquery} between 2026 July and September. Of the
\ncutmax{} cutouts in the SPT-CL~J0546$-$5345 field, \ndeepcut{} are drawn
from the \texttt{spherex\_qr2\_deep} collection. All cutouts in the other
fields are drawn from \texttt{spherex\_qr2}. Each cutout is a square whose
side equals twice the $R_{200}$ of the galaxy cluster \citep{bahk2026clusters}
plus a margin for the PSF, where $R_{200}$ is the radius within which the
mean density is 200 times the critical density. For the Coma cluster, whose
angular extent is large, the side equals $R_{200}$; for the COSMOS field, it
equals the diameter of the catalog footprint. The cutouts range from
\cutboxmin{} to \cutboxmax{}
native pixels on a side (\cutarcmin{}). Each field contributes between
\ncutmin{} and \ncutmax{} cutouts depending on its visit coverage,
for a total of \ncuttotal{} cutouts.

Each Level~2 file holds the image, its variance, a flag plane, the
zodiacal-light model, and per-pixel wavelength and solid-angle maps. Each
file also carries the exposure-averaged, oversampled optical PSF cube of
the detector, with one plane per detector zone
\citep{akeson2025pipeline,korngut2026instrument}. We use the QR2 calibration
products throughout, without later wavelength or gain corrections,
and apply the QR2 erratum on the
ordering of the planes of the cube \citep{irsa2025qr2}. Source positions, morphological
types, and profile parameters come from \ls{} \citep{dey2019legacysurveys};
the unWISE W1 and W2 forced photometry of the same catalog
\citep{lang2014unwise,schlafly2019unwise} serves the flux-domain comparisons
of Section~\ref{sec:photoz}.

\subsection{The Comparison Sample}\label{sec:cohort}

We compare all configurations on the same fixed sample of galaxies with
spectroscopic redshifts, so that a difference between two fits reflects
the photometry rather than a change in sample composition. We match the
\ls{} sources within $1\arcsec$ to the spectroscopic compilation of
\citet{bahk2026clusters}, which combines SDSS DR17 \citep{abdurrouf2022sdssdr17}
supplemented by \citet{hwang2010lirgs,hwang2014clusters}, DESI DR1
\citep{desi2025dr1}, and galaxy-cluster spectroscopy from SPT
\citep{bleem2015sptsz}, ACT \citep{sifon2016act}, and GOGREEN
\citep{balogh2021gogreen}. In the COSMOS field, it
also includes the COSMOS spectroscopic compilation
\citep{khostovan2025specz}, matched through COSMOS2025
\citep{shuntov2025cosmos2025}.
A source enters
the sample if it has $0<\zspec<3$, an extended \ls{} morphology (any type
other than \texttt{PSF}), and visit coverage at least equal to that of the
field core. Coverage is the number of distinct cutouts that contain the
source; the core value is the median coverage within $3\arcmin$ of the field
center. Every member is therefore observed at the representative
coverage of the field.

Throughout the paper, we divide sources at the \ls{} $z$-band magnitude
$\mz=21$, where $\mz=22.5-2.5\log_{10}f_z$ and $f_z$ is in nanomaggies.
This is the depth at which a source stops being individually detectable in
its coadded \spherex{} spectrum. Toward fainter magnitudes, the S/N in a
single channel falls to 1 at $\mz\simeq\snonecross$, whereas that of the
inverse-variance-weighted mean flux over all 102 channels falls to
S/N $=3$ only at $\mz\simeq\snthreecross$. This second
crossing depends on visit coverage and ranges over \snthreerange{} across our
fields; we adopt \zbound{}, near its pooled value, as one boundary for all
fields rather than as an optimum. Section~\ref{sec:identifiability} tests
other positions.

The \zcut{} comparison sample contains $N_{\rm comp}=\nexternal{}$
sources, of which the COSMOS field supplies \ncosmosext{} (\cosmosshare{})
and the Coma field \ncomaext{} (\comashare{}).
The Coma field also dominates the low-redshift end, with \ncomalowz{} of the
\nlowzext{} sources at $\zspec<0.1$. Because two fields hold most of the
sample, we report every pooled difference between configurations together
with an equal-field counterpart, in which each field carries the same
weight (Appendix~\ref{app:fields}).

\section{Methods}\label{sec:method}

\subsection{\tractorjax{}: Forward Model and Implementation}\label{sec:model}

Because positions, morphologies, and the PSF are frozen at their catalog and
calibration values, the model for a cutout is linear in the source
fluxes and is fitted by weighted least squares. The model is
\begin{equation}
 \boldsymbol d=A\boldsymbol f+B\boldsymbol b+\boldsymbol\epsilon,
 \qquad \boldsymbol\epsilon\sim\mathcal N(0,W^{-1}),
 \label{eq:forward}
\end{equation}
where $\boldsymbol d$ holds the pixels, and $\boldsymbol\epsilon$ is Gaussian
noise with zero mean and covariance $W^{-1}$, in which $W$ is the diagonal
inverse-variance weight matrix. Each column of $A$ is the unit-flux,
PSF-convolved image of one cataloged source; each column of $B$ is
the indicator of one tile of the cutout, so that $\boldsymbol b$ holds
one background constant per tile. With $M=[A\;B]$ and $\boldsymbol\theta=(\boldsymbol f,\boldsymbol b)$,
the solution satisfies the normal equations $G\boldsymbol\theta=\boldsymbol h$
with $G=M^{\top}WM$ and $\boldsymbol h=M^{\top}W\boldsymbol d$. The pixels
undersample the PSF in every band, whereas the reference catalog fixes
positions and shapes far more precisely than these images could. A free
fit would therefore spend its freedom on quantities that the data do not constrain.

Each cutout is prepared once before the fit. Pixels carrying any of the
flag bits \texttt{SUR\_ERROR}, \texttt{PHANMISS}, \texttt{NONFUNC},
\texttt{MISSING\_DATA}, \texttt{HOT}, \texttt{COLD}, \texttt{PERSIST}, or
\texttt{OUTLIER} \citep[Table~16]{spherex2026expsupp} receive zero weight.
These pixels and those carrying the \texttt{SOURCE} bit are excluded from the
background estimate. We
subtract the Level~2 zodiacal-light model plus two background terms
fitted on the remaining pixels, one in wavelength and one in position.
The wavelength term comes first because a linear variable filter maps
wavelength onto detector position. A background feature at a fixed
wavelength, above all the He~\textsc{i} \helinewave{}\,\um{} airglow
line, therefore appears as a sharp-edged stripe that a spatial estimate cannot follow. The
sharp stripe becomes a smooth background profile when it is plotted
against the wavelength assigned to each pixel. Fitting that profile
removes the stripe along with any other wavelength-dependent component; the spatial term then fits the remainder. The data and
variance are converted from surface brightness to mJy per pixel with the
per-pixel solid angle. The background vector $\boldsymbol b$ in Equation~\ref{eq:forward}
absorbs what remains in each tile.

Every source is rendered on a grid oversampled five times relative to the
$6\farcs15$ native pixel and integrated back onto the native pixels with
its flux conserved. Point sources are the oversampled PSF shifted to the
catalog position; galaxies are their \ls{} S\'ersic profiles, represented
as Gaussian mixtures \citep{hogglang2013mixture}, convolved with the same
PSF in Fourier space. The PSF of a tile is interpolated bilinearly at the
tile center from the four surrounding zone planes of the Level~2
cube, each recentered on its measured core (an offset of
$\sim$\psfcoreoffset{} native pixel, measured once for each detector zone).

\tractorjax{} reimplements the catalog-forced mode of \thetractor{}
\citep{lang2016tractor} in \jax{} \citep{jax2018github}. It runs on GPUs in
double or single precision. Each cutout is divided into tiles, each a square core
plus a halo that includes the neighboring sources whose PSF wings extend
into the core. The flux of each source is taken from the tile whose core
contains it. The core
and halo widths are adjustable; this paper uses \tilecore{} and
\tilehalo{} pixels. The source arrays of each cutout are padded to a few
fixed sizes so that one compiled kernel serves many cutouts; the CPU
prepares the next cutout while the GPU fits the current one. The package
\tjxspherex{} forms the \spherex{} layer outside the engine and handles
catalog ingestion, WCS projection, PSF and wavelength preparation, and the
assembly of the 102-channel spectra.

Every rate quoted in this paper
was measured with \jax{} \jaxversion{} on its CUDA~12 build. The timing
set consists of \ntimingcut{} cutouts of the A2537 field, each
$\timingpix\times\timingpix$ native pixels, from a retrieval that was
earlier and smaller than the $R_{200}$ cutouts of Section~\ref{sec:images}.
These cutouts hold a median of \timingnsrc{} sources each
at full depth and \timingnsrczcut{} at \zcut{}; the CPU reference
cycles through the same cutouts on one core.

\subsection{Fitted-Catalog Depth and Regularization}\label{sec:configs}

We fit five configurations, each pairing a fitted catalog with a
choice of regularization (Table~\ref{tab:estimators}). The reference catalog truncated at
\zbound{} and the full catalog are each fitted without regularization
(\linearest{}) and with a Jacobi-normalized eigenvalue floor (\efl{}).
The full catalog is also fitted with the floor and with external flux
priors on the sources fainter than the boundary (\eflp{}). Sparse selection
(\lassoest{}) in place of truncation is defined in
Appendix~\ref{app:lasso}.
Every configuration solves the normal equations of
Section~\ref{sec:model}; they differ in how $G$ is conditioned before it
is inverted.

\providecommand{\rg}{\raggedright\let\\\tabularnewline}
\begin{deluxetable}{llp{0.42\columnwidth}}
\tabletypesize{\footnotesize}
\tablecaption{The five configurations fitted: the catalog each fits and
the regularization that makes the fit well posed. \label{tab:estimators}}
\tablehead{\colhead{Configuration} & \colhead{Fitted catalog} &
\colhead{Regularization}}
\startdata
\rg \linearest{} & \rg to \zbound{} & \rg none \\
\rg \efl{} & \rg to \zbound{} & \rg eigenvalue floor \\
\rg \linearest{} & \rg full & \rg none \\
\rg \efl{} & \rg full & \rg eigenvalue floor \\
\rg \eflp{} & \rg full & \rg eigenvalue floor; Gaussian flux priors on sources fainter than \zbound{} \\
\enddata
\tablecomments{The eigenvalue floor is Jacobi-normalized at
$\eigfloorfrac\lambda_{\max}$ (Section~\ref{sec:configs}). The priors are Gaussian,
centered on expected fluxes from the \ls{} optical and WISE photometry, with width
$\max(\priorsigmafrac\,f, \priorsigmamin)$; targets brighter than \zbound{}
carry none. Sparse selection is defined in Appendix~\ref{app:lasso}.}
\end{deluxetable}

For \efl{}, let $D=\operatorname{diag}(G)$ and
$\widehat G=D^{-1/2}GD^{-1/2}$; each eigenvalue $\lambda_k$ of
$\widehat G$ is replaced by $\max(\lambda_k,\eigfloorfrac\lambda_{\max})$
before inversion \citep[cf.\ the covariance-matrix floor
of][Appendix~C7]{behroozi2019universemachine}. Because the normalization gives
every column unit weight, neither the background nor a bright source
sets the scale. The floor then
acts only on the directions of $\widehat G$ with eigenvalues near zero:
the anticorrelated flux splits between sources that
the pixels do not separate. A source that the images constrain on its own
sits at an eigenvalue near one and is solved exactly. The floor imposes neither
positivity nor the removal of any source; every fitted source keeps a
flux.

The prior configuration adds a Gaussian prior on the flux of every source
fainter than \zbound{}, the \emph{nuisance} population, and
none directly on the brighter \emph{protected} targets. Before eigenvalue
flooring, the normal equations are
$(G+\Lambda)\boldsymbol\theta=\boldsymbol h+\Lambda\boldsymbol\theta_{\rm
prior}$, where $\Lambda$ is diagonal with $\Lambda_{jj}=\sigma_j^{-2}$ for
a nuisance source and zero for protected targets and background
coordinates. For numerical conditioning, each normalized prior precision
$\Lambda_{jj}/G_{jj}$ is capped at \priorprecisioncap{}.
The floor is applied to the normalized $G+\Lambda$, with the same $D$
and threshold $\eigfloorfrac\lambda_{\max}(\widehat G)$ as the fit
without priors; both are computed from $G$ alone.
The prior of a source is centered on its expected flux $f_{\rm prior}$.
We obtain $f_{\rm prior}$ by interpolating the \ls{} $g$, $r$,
$i$, $z$ and unWISE W1, W2 photometry linearly in $\ln f$
against $\ln\lambda$ to the wavelength of the source in each cutout.
Beyond the outer bands, the end slope is extrapolated under
$|d\ln f/d\ln\lambda|\le\priorslopemax{}$. Its width is
$\sigma_j=\max(\priorsigmafrac\,f_{\rm prior},\priorsigmamin)$.
The minimum width prevents very faint flux predictions from imposing
arbitrarily tight priors.

The reported covariance has a different meaning in each configuration.
For \linearest{}, $G^{-1}$ is the sampling covariance of the fit, the
pixel noise propagated through Equation~\ref{eq:forward} with positions,
shapes, the PSF, and catalog membership taken as exact. For \efl{}, the
floored inverse is not the covariance of the estimate. In a direction that the
floor damps, the fit is pulled toward zero by a factor that the inverse does
not record; the scatter in that direction is smaller than the
$1/(\eigfloorfrac\lambda_{\max})$ that the inverse reports. The covariance of
the estimate is $\widehat G_{\rm f}^{-1}\widehat G\,\widehat G_{\rm f}^{-1}$,
where $\widehat G_{\rm f}$ is the floored matrix, but it omits this bias
toward zero and vanishes as the eigenvalue approaches zero. We therefore
report the floored inverse, a conservative uncertainty that remains large for
fluxes that the data do not constrain. For
\eflp{}, $(G+\Lambda)^{-1}$ is the Gaussian posterior covariance before
eigenvalue flooring, where $\Lambda$ is the capped prior precision.
The reported variances are the diagonal elements of the floored inverse and equal
the posterior variances when the floor is inactive. When the floor is
active, they include additional numerical regularization.

\subsection{Photometric Redshifts, Broadband Comparison, and Paired Contrasts}\label{sec:photoz}

We derive photometric redshifts for each configuration in three steps:
assembling the 102-channel spectra, fitting them with galaxy templates, and
calibrating the fits. We assemble the visit-level measurements into 102-channel spectra by
inverse-variance averaging the visits that fall in each channel. Every
configuration is fitted with \eazy{} \citep{brammer2008eazy} on the
\ntemplates{} galaxy templates that \citet{feder2024spherex} compiled from
the Brown atlas \citep{brown2014atlas} and the COSMOS templates
\citep{ilbert2009cosmos}, the library that \citet{bae2026redshifts} and
\citet{bahk2026clusters} used for \spherex{}. The channel
responses are Gaussians
built from published band parameters alone. Each response is centered on the pivot
wavelength recorded in the Level~2 header; its FWHM is $\lambda_{\rm c}/R$, where $R$ is the resolving power of the band plan
\citep{bock2025spherex,korngut2026instrument}.
Appendix~\ref{app:lab-response} repeats the redshift analysis with the
laboratory-measured responses \citep{hui2026spectral}. Each configuration then
carries its own calibration chain, recorded in Appendix~\ref{app:fields}.
The chain sets channel zeropoints at fixed \zspec{}, applies a
simulation-derived S/N weighting, and includes template-error terms whose
amplitude is chosen by the same rule for all configurations. A
smooth warp then maps the raw template-fit redshift $z_{\rm raw}$ to the
final point estimate \zphot{}.

For every source with a calibrated redshift, we define
\begin{equation}
 \begin{aligned}
 \dz &\equiv \frac{\zphot-\zspec}{1+\zspec},\\
 \signmad &\equiv 1.4826\,\operatorname{MAD}(\dz),\\
 \operatorname{MAD}(x)&\equiv
 \operatorname{median}|x-\operatorname{median}(x)|,
 \end{aligned}
 \label{eq:photoz-metrics}
\end{equation}
and the catastrophic fraction $\etaout=P(|\dz|>0.15)$, the conventional
definitions \citep{ilbert2006photoz}.

We apply two selections: one that keeps spectra with enough signal to
calibrate the redshifts, and one that defines the reliable sample for the
science comparisons. The calibration pool of a configuration
contains the spectroscopic sources brighter than \zbound{} that are used to
calibrate its redshifts. A source enters the pool only if its median channel \snpresel{}. The science comparisons use the reliability criterion
of \citet{bahk2026clusters},
\begin{equation}
 \crit:\quad N_{\rm ch}(\mathrm{S/N}>5)>50,
 \label{eq:criterion}
\end{equation}
where $N_{\rm ch}(\mathrm{S/N}>5)$ is the number of the 102
channels in which a source is detected at S/N above 5. The criterion
therefore requires a detection in at least half of them. \ncrit{} and \fcrit{} are the number and fraction of
sources in the comparison sample that satisfy \crit{} in a given
configuration. We confirm that every source satisfying \crit{} also
enters the calibration pool and therefore has a calibrated redshift.
For comparisons at matched accuracy, we raise a threshold on
$N_{\rm ch}(\mathrm{S/N}>3)$, the number of channels detected at S/N above 3,
which removes the least constrained spectra first. The S/N level is lower
than in \crit{} so that the threshold also ranks fainter sources, which rarely
reach S/N $=5$ in many channels. At each threshold, we
record the sample size and \signmad{}; for a given \signmad{}, we keep the
largest sample whose scatter does not exceed it.

Because photometric redshifts are insensitive to an overall flux scale,
we also test the fluxes directly against broadband catalog photometry. We
pass the spectrum of each configuration through the \ls{} $z$ and the
unWISE W1 and W2 bands and compare the resulting synthetic fluxes with the
catalog photometry of the same source in those bands, hereafter the
\emph{broadband comparison}. Each band is summarized by
\begin{equation}
 F_{\mathrm{\spherex}} = g\,F_{\rm LS/WISE} + c,
 \label{eq:broadband}
\end{equation}
where $g$ is the multiplicative response and $c$ is the additive offset.
We estimate both with a Theil--Sen fit on the extended sources brighter
than \zbound{} that carry a flux in every configuration. Because each flux
measurement is affected by the depth of its own fitted catalog, $c$
represents an offset between the two measurements rather than an absolute
flux error.

We compare these photometric redshifts and synthetic fluxes between
configurations source by source and assess the significance of each
difference. Each comparison is a paired contrast: the difference from the
\zcut{} \efl{} fit on the sources that both configurations measure. Its
uncertainty comes from a
bootstrap that resamples sources within each field (\nboot{} draws).
Throughout the paper, intervals in square brackets give the central
\pairedconfidence{} of the bootstrap draws unless stated otherwise. For
catastrophic outliers, we count the sources that are catastrophic in one
configuration but not in the other, in both directions, and test the two
counts against each other with McNemar's exact test
\citep{mcnemar1947test}. We repeat every
contrast with each field given equal weight (Section~\ref{sec:cohort})
and with one field left out at a time. Because each configuration is
calibrated on its own, a contrast includes the difference between the two
calibrations as well as the difference between the two photometries. We report
every interval and every pair of these counts without correcting for
multiple comparisons.

\subsection{Truth-Controlled Simulations}\label{sec:sims}

Every truth-controlled test in this paper uses simulated scenes. We
build each scene from two ingredients: a real QR2 cutout of A2537 and, for every \ls{} source in the
field, a model flux at the cutout wavelength that serves as the simulation
truth.
Because the cutout supplies the WCS, the delivered PSF cube, and the flag
and variance planes, the noise in the scene is that of the observed data.
We give each source the template
from the library of Section~\ref{sec:photoz} and the redshift that
together best match its own \ls{} $r-z$ and $z-$W1 colors. We then integrate
that template through the channel response and scale it to the
catalog $z$-band flux of the source. A template-free interpolation between the
$z$ and W1 fluxes of the source reproduces the result to a median ratio of
\truthchordratio{} on the sources brighter than $\mz=\truthbrightmag{}$.

Three experiments use these scenes. The depth experiment
(Section~\ref{sec:identifiability}) takes one detector-3 cutout at
\depthwave{}\,\um{} and renders the scene with the forward model of the fit.
It then draws \nrealiz{} noise realizations and fits the catalog truncated at
each of several boundaries in $\mz=\plateaurange$ and at full depth. The
engine
check (Section~\ref{sec:engine}) takes \npullcut{} cutouts, two per
detector over \pullwaverange{}\,\um{}. It renders them by an independent path,
\galsim{} \citep{rowe2015galsim} with exact S\'ersic profiles and the
delivered PSF, so that the fit is not tested against its own forward
model. The prior
check (Section~\ref{sec:engine}) fits the same cutouts with the prior
configuration exactly as run on the observational data. We test four
cases: exact priors, priors wrong by a lognormal scatter of \priorwrongsig{},
no priors, and the catalog-based priors of Section~\ref{sec:configs}.

The depth experiment measures how many of the cataloged sources a
cutout can constrain, using two statistics: the flux uncertainty each source
would have in isolation and the effective number of degrees of freedom that a
fit spends on the pixels. A source is \emph{measurable in isolation} when
$f_{\rm true}>\sigma_{\rm iso}$, where
\begin{equation}
 \sigma_{{\rm iso},j}=(a_j^{\top}Wa_j)^{-1/2}=G_{jj}^{-1/2}
 \label{eq:sigiso}
\end{equation}
is the uncertainty of a fit to source $j$ alone, with $a_j$ its column
of $A$ and $W$ the inverse variance of Equation~\ref{eq:forward}. Because $\sigma_{\rm iso}$
is the noise of a PSF fit to an isolated source, the criterion
requires S/N above 1 in the absence of neighbors.
In an unregularized fit, a neighbor that shares its pixels can only
widen this uncertainty. The effective number of degrees of freedom is
$\operatorname{tr}(H)$, where
\begin{equation}
 \hat{\boldsymbol m}=H\boldsymbol d,\qquad
 H=M\,G_{\rm f}^{-1}M^{\top}W
 \label{eq:hat}
\end{equation}
maps the data pixels to the fitted model pixels, with
$G_{\rm f}^{-1}=D^{-1/2}\widehat G_{\rm f}^{-1}D^{-1/2}$ the floored inverse
of Section~\ref{sec:configs}, which equals $G^{-1}$ when no eigenvalue is
floored. For
\linearest{}, $\operatorname{tr}(H)$ equals the number of fitted
parameters. Under the floor, each lifted mode contributes the ratio of
its eigenvalue to the floor instead of one. Because both statistics depend only
on the catalog, the PSF, and the variance plane, they take the same
value in every noise realization.

The simulated scenes have three limitations. They cover one field; the
depth experiment uses one single-band cutout, whose densest region is the
galaxy-cluster core around the brightest cluster galaxy (BCG). The depth scene shares
its forward model between rendering and fitting. The absolute flux scale of
a scene is an extrapolation of the optical and WISE photometry of each source to
the \spherex{} wavelength and is uncertain at the level of tens of percent.
We therefore quote ratios of flux scatter between depths.

\section{Results}\label{sec:results}

Here we test the engine and then use it to measure the cost
of fitting past what the images resolve. We first verify the
fluxes and uncertainties that \tractorjax{} reports and measure how fast it
runs (Section~\ref{sec:engine}). With depth as the variable, we then
measure against truth how many source fluxes the images determine
(Section~\ref{sec:identifiability}). Using observational data, we
evaluate two approaches for extending the fit beyond that limit:
regularization alone (Section~\ref{sec:retention}) and external flux
priors (Section~\ref{sec:prior}).

\subsection{\tractorjax{}: Verified Uncertainties and Performance}\label{sec:engine}

\begin{figure*}
\centering
\includegraphics[width=\textwidth]{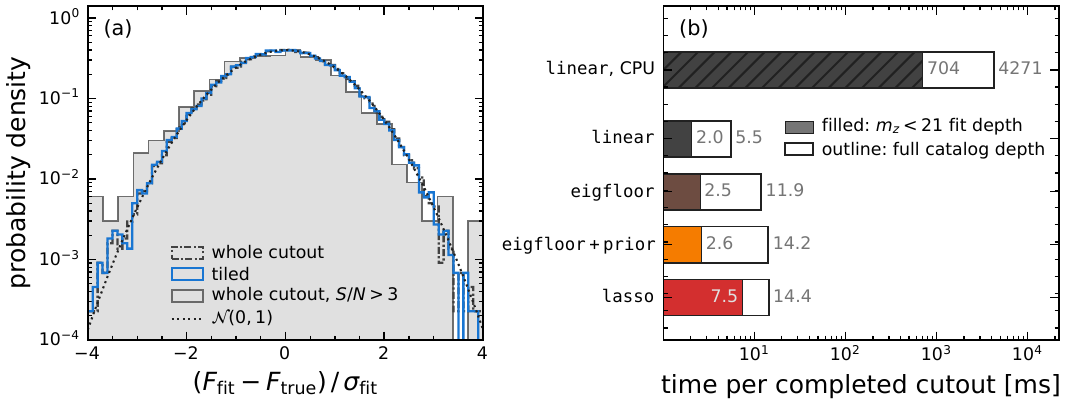}
\caption{Flux uncertainty tests and timing comparisons for \tractorjax{}.
(a) Pull distribution of the plain fit on the independently rendered
scenes of Section~\ref{sec:sims}, \npullmeas{} measurements of \npullsrc{}
sources; shaded, the \npullsnrthree{} measurements above S/N $=3$, where
the forward model can be tested. (b) Time per completed cutout for
\thetractor{} on one CPU core (hatched) and \tractorjax{} on one NVIDIA
H100 of the same node, same day, \ntimingcut{} A2537 cutouts of
$\timingpix\times\timingpix$ pixels. GPU values are the best measured service
rates, several cutouts co-scheduled, not single-cutout latencies. The full-depth \linearest{} row is timed in
single precision, a timing device, not the double-precision fit of
Table~\ref{tab:estimators}.
\label{fig:engine}}
\end{figure*}

The uncertainties that \tractorjax{} \citep[v0.3.0;][]{bahk2026tractorjax}
reports match the scatter of its errors against truth. We test them with the pull, the normalized residual
$(F_{\rm fit}-F_{\rm true})/\sigma_{\rm fit}$, whose distribution is a
unit Gaussian for an unbiased linear fit with Gaussian noise and correct
uncertainties. On the
independently rendered scenes of Section~\ref{sec:sims}, the pull of the
plain fit has a width \signmad{} of \pullwhole{} in the whole-cutout
geometry and \pulltiled{} in the tiled geometry. These widths are
measured over \npullmeas{}
measurements of \npullsrc{} sources (left panel of Figure~\ref{fig:engine}). The
population is the full \ls{}
catalog at the depth fitted in the observational data and is faint by nature; \pullbelowthree{} of the
measurements lie below S/N $=3$, where noise makes the pull less sensitive to
forward-model mismatch. On the \npullsnrthree{} measurements
above S/N $=3$, the width is \pullsnrthree{}; the reported errors are therefore
within \pullsnrthreepct{} of the delivered scatter.

The galaxy model has one known limitation. When the true profile is an
exact S\'ersic profile, as in these scenes, the truncated Gaussian mixture
underestimates its flux by \sersicwingterm{} at S/N $>3$, because the
mixture omits the outer wings. Point sources are recovered without bias,
with a median flux ratio of \pullpoint{}. We keep the mixture in the
photometry because real galaxies are not exact S\'ersic profiles and
because the \ls{} catalog itself was fitted with the same mixture. The wing
correction of \citet[Appendix~B]{bahk2026clusters} reduces the term to
\sersiccorrresid{}.

We test the reported uncertainties of the prior configuration on the same
scenes with priors that are exact, wrong, absent, or catalog-based.
The test uses the protected targets, the sources brighter than \zbound{}
that carry no prior. Over \npullprior{} measurements, the pull width is \pullpriorexact{} with exact priors,
\pullprior{} with priors wrong by a lognormal scatter of \priorwrongsig{},
\pullpriornone{} with no prior, and \pullpriorls{} with the catalog-based
prior of Section~\ref{sec:configs}. The catalog-based prior differs from the
simulation truth by \lspriornmad{}\,dex in NMAD, with a mean offset of
\lspriormean{}\,dex. The widths differ by at most \pullpriormaxdiff{}
(95\% interval \pullpriormaxdiffci{}); the residual scatter relative
to the reported target uncertainties therefore changes little across these prior
choices.

\tractorjax{} reproduces the fluxes of \thetractor{}. On an identical
\zcut{} problem of \nparity{} sources in an observed A2537 cutout, the two
agree to a median flux ratio within \parityaltmed{} of unity. The regularized
fits run in single precision, in which the BCG flux in that cutout stays
within \fptwoeigrel{} of its double-precision value. In contrast, the
unregularized full-depth fit is solved in double precision, because
single-precision roundoff nearly triples the BCG flux in this cutout.

A fit takes milliseconds on one GPU, whereas the reference implementation
takes seconds on one CPU core. When several cutouts are processed
concurrently on the GPU (co-scheduling), one NVIDIA H100 completes a full-depth
cutout in \eigservicems{}\,ms and a \zcut{} cutout in \hzcuteigms{}\,ms
(right panel of Figure~\ref{fig:engine}). The reference implementation (\thetractor{}
through the \texttt{cpu-tractor} backend of \tjxspherex{}) takes
\cpucoresec{}\,s per full-depth cutout and \cpucoreseczcut{}\,s per \zcut{}
cutout on one CPU core. Because these are service
rates on one field, the ratio between them shows
how many CPU cores one GPU replaces, not how much sooner a single cutout
finishes.

The comparison is conservative for the GPU and fair to the reference
implementation. Both H100 rates are limited not by the GPU but by the
\lampjobcores{} host CPU cores available to the benchmark; the full-depth rate
changed by \hloadspread{} between sessions with the varying load from other
users. With more host
cores per GPU, the rates would rise; they are a lower limit
on the capability of the GPU. The reference values use the standard
forced-photometry path of \thetractor{}, in which \cpurenderfrac{} of
the time goes to rendering the per-source models in compiled C; no
choice of solver would change them much. Run in the CPU mode of \jax{},
\tractorjax{} is slower than \thetractor{}, because its batched dense
design is intended for GPUs; the comparison is therefore between each
implementation on the hardware for which it is built.

The lower-cost NVIDIA L40S delivers comparable performance, with a
substantial gap relative to the H100 only for full-depth fits. It
requires \lfortyeigservicems{}\,ms per fit at full depth and
\lfortyzcuteigms{}\,ms at \zcut{}. On the L40S, the GPU itself is the
limit; it is fully occupied at full depth by a single stream and at
\zcut{} under the best co-scheduling setting. The
L40S is \cardgapfull{} times slower than the H100 at full depth but only
\cardgapzcut{} times at \zcut{}, where the L40S is nearly as good as the H100.
Because the H100 rates are limited by the host CPU cores whereas the L40S
is limited by the GPU itself, these ratios are lower limits on the gap between
the two GPUs.

At these rates, a re-photometry of the \missionimages{} spectral
images of the mission (about \missioncutouts{} cutout fits) would take about
\missioncarddays{} GPU-days at full depth and \missioncarddayszcut{}
GPU-days at \zcut{} on one H100. The reference implementation would take
about \missioncoreyr{} core-years at full depth and \missioncoreyrzcut{}
core-years at \zcut{}. These estimates
assume \cutoutsperimage{} cutouts of $\timingpix\times\timingpix$ pixels
per $\activepix\times\activepix$-pixel spectral image, six per exposure,
the measured tiling, and the A2537 source density as an upper bound.
The rates include reading the cutout files but exclude downloading the
images and saving the output catalogs. Because a day of survey data costs
\dailycardhours{} GPU-hours, one GPU outruns the data stream; the \zcut{}
estimate holds almost unchanged on the L40S.

\subsection{The Number of Source Fluxes Constrained by the Images}\label{sec:identifiability}

\begin{figure}
\centering
\includegraphics[width=\columnwidth]{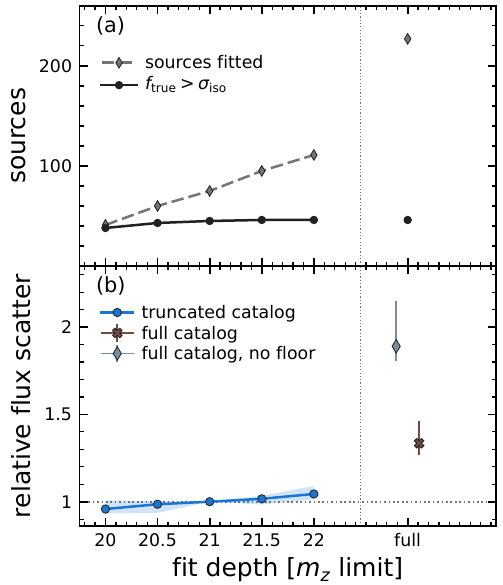}
\caption{Source counts and relative flux scatter across fitted catalog
depths for \nrealiz{} noise realizations of a simulated A2537 cutout at
\depthwave{}\,\um{}.
(a) Counts of fitted sources and those measurable in isolation
($f_{\rm true}>\sigma_{\rm iso}$) within \depthapradius{} pixels of the BCG.
(b) Scatter of fractional residuals for the common sources,
relative to the \zcut{} fit of the same realization. Points mark median ratios;
shaded regions and error bars span the \depthpercentiles{} percentiles of
these ratios across noise realizations.
\label{fig:identifiability}}
\end{figure}

The images run out of individually measurable sources well before the
reference catalog is exhausted
(top panel of Figure~\ref{fig:identifiability}). Within \depthapradius{} pixels of
the A2537 BCG, the densest region of the simulated
scene, opening the catalog from \zcut{} to full depth raises the number
of fitted sources from \cutfitn{} to \fullfitn{}. The number of sources
measurable in isolation (Equation~\ref{eq:sigiso}) rises only from \cutmeasn{} to
\fullmeasn{}; the measurable fraction therefore falls from \cutmeasfrac{} to
\fullmeasfrac{}. Nearly every added source is fainter than its
$\sigma_{\rm iso}$ but still enters the fit as a free parameter.
Over the whole cutout, the full-depth fit has \deptheffratio{} times
the effective number of degrees of freedom of the \zcut{} fit
(Equation~\ref{eq:hat}). Most of the added freedom therefore goes to
sources too faint to constrain, whereas the number of individually
measurable sources is nearly unchanged.

Fitting these additional sources increases the flux scatter even for sources
that the truncated catalog already contains
(bottom panel of Figure~\ref{fig:identifiability}). Because we compare the same
\popncommon{} sources fitted at every depth, changes in sample
composition cannot account for the difference.
For each noise realization, we divide the \depthpercentiles{} percentile
width of $(f-f_{\rm true})/f_{\rm true}$ across these sources by the width
from the \zcut{} fit of the same realization. The median of these \nrealiz{}
ratios is \popscatterratio{} for full-depth \efl{}, with a
\depthpercentiles{} percentile range of \popscatterci{}.

Regularization decreases the scatter, but the full-depth fit still has
larger scatter than the \zcut{} fit.
Without the floor, the full-depth fit has a median scatter ratio of
\popscatterlinratio{} relative to the same \zcut{} reference.
In the full-catalog matrix, the floor lifts
\modesliftfull{} eigenmodes, suppressing combinations of neighbor fluxes
that the image constrains only weakly.
At \zcut{}, by contrast, the floor changes only \modesliftcut{} modes;
the median fractional flux difference from the unregularized fit is
\cuteigagree{}. The smaller scatter of the \zcut{} fit therefore comes
from truncation itself, not from the floor.

The presence of a truncation boundary matters more than its precise position.
Across the truncated depths in $\mz=\plateaurange$, the median paired
scatter ratios relative to \zcut{} span only \scatterrunglo{} to
\scatterrunghi{} (bottom panel of Figure~\ref{fig:identifiability}). Moving the boundary
from \zbound{} to $\mz=\depthfaintbound{}$ produces a small increase,
with the wider distribution in \scatterfaintworse{} of \nrealiz{} draws.
However, the full-depth fit has greater scatter than
the fit at every truncated depth in every realization for both estimators.
\citet{huai2026confusion} likewise restrict the fitted catalog for their
cosmology sample, selecting $\mz<\huaizcut{}$ plus fainter sources passing
a $z-W1$ color cut.

\subsection{Regularization Alone: Fewer Sources at a Given Redshift Accuracy}\label{sec:retention}

We now test whether regularization alone can preserve the
photometric-redshift sample when the fitted catalog is extended to full
depth. Using the observational data, we compare the \zcut{} and full-depth
\efl{} configurations on the sample defined in Section~\ref{sec:cohort},
without external flux priors. We assess both the sample size at a given
photometric-redshift scatter and the scatter on the sources that satisfy \crit{}
(Equation~\ref{eq:criterion}) at both depths.

\begin{figure*}
\centering
\includegraphics[width=\textwidth]{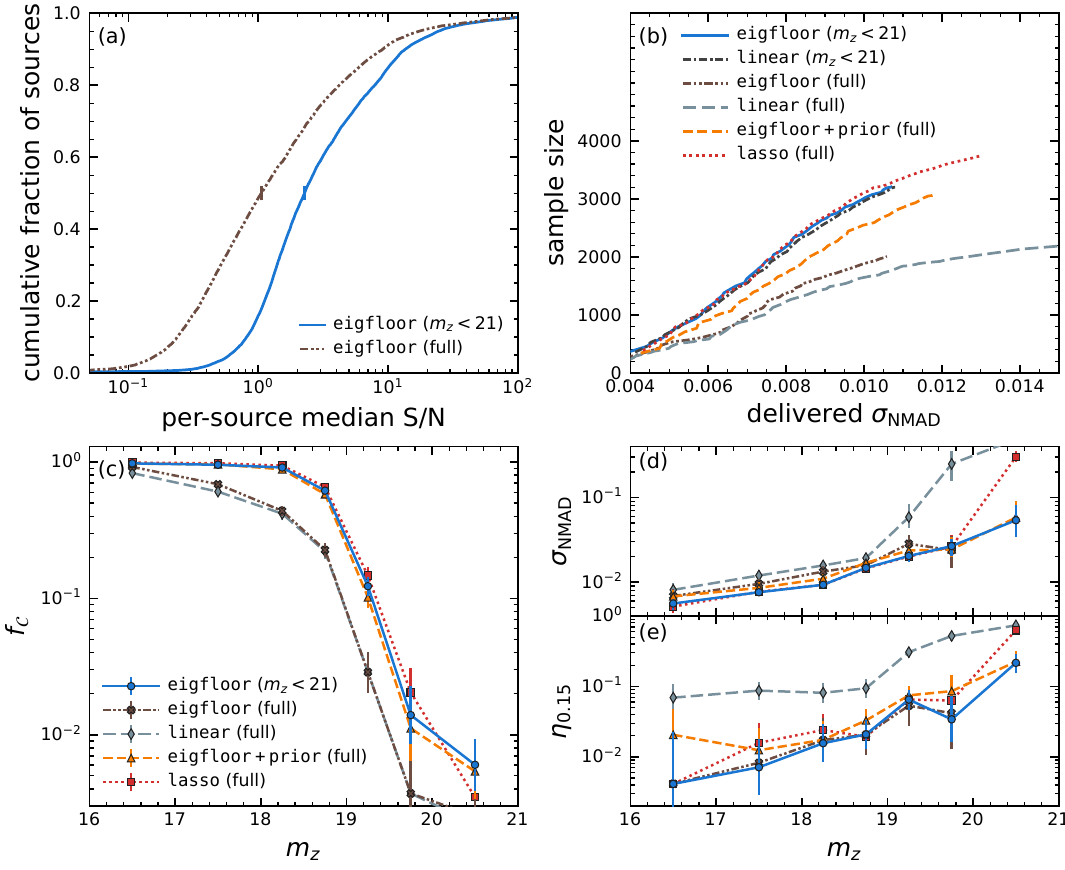}
\caption{Spectral S/N, sample size, and photometric-redshift accuracy for
different fitted catalog depths and estimators.
(a) Cumulative distribution of per-source
median S/N on the comparison sample; ticks mark the medians at each depth.
(b) Largest sample reaching a given \signmad{}, from selections of
increasing strictness on $N_{\rm ch}(\mathrm{S/N}>3)$, scored with the
calibrated redshifts of each configuration. (c) Fraction \fcrit{} of the comparison sample
satisfying \crit{} (Equation~\ref{eq:criterion}) vs. catalog
magnitude. (d, e) \signmad{} and \etaout{} vs. magnitude, on
the calibrated sources of each configuration. These curves compare
different samples and do not establish the paired accuracy difference
(see text). The faint bin of the sparse fit reflects the
selection effect described in Appendix~\ref{app:lasso}.
\label{fig:retention}}
\end{figure*}

Because the \efl{} fit assigns large uncertainties to the fluxes that the
data do not constrain, extending the fitted catalog to full depth shifts the
whole distribution of per-source median S/N downward, from a median of
\sciencemedsnr{} to \fullmedsnr{} (top left panel of Figure~\ref{fig:retention}).
A reliability requirement converts this shift into a smaller sample.

At a given required photometric-redshift scatter, the full-depth fit
supplies fewer sources. The \zcut{} fit yields \amratiolo{}--\amratiohi{} times as many sources as the full-depth
fit across $\signmad=\amsigmalo$--\amsigmahi{}
(top right panel of Figure~\ref{fig:retention}). At lower scatter, the
ratio is smaller but remains above unity. The \zcut{} fit also yields more
sources at matched scatter when sources are selected by the odds parameter
of \citet{hernancaballero2021minijpas}, the posterior probability within a
prescribed interval around the best-fit redshift.

Fewer sources satisfy the reliability criterion \crit{} in the full-depth
fit: \ntierfull{}, against \ntier{} in the \zcut{} fit (Table~\ref{tab:retention}).
The two selected samples have similar redshift scatter
($\signmad=\tierdeliv$ at \zcut{} and \tierfulldeliv{} at full depth), but the scatter
is larger at full depth when we compare the sources that satisfy \crit{} at both depths. Of the net loss of
\ncritlost{} sources, \critlostfrac{} falls in
$\mz=\critlostmaglo$--\pairedbinmax{}
(bottom left panel of Figure~\ref{fig:retention}). The full-depth sample is brighter, with
median $\mz=\medmzcritfull$ against \medmzcrit{} for the truncated fit.
The sources that satisfy \crit{} only in the truncated fit have median $\mz=\lostmedmz$
and $\signmad=\lostnmadcut$ in that fit. Removing these fainter, noisier
sources masks part of the accuracy loss; on the \npairedcrit{} sources
meeting \crit{} at both depths, the full-depth \signmad{} is larger by
a factor of $\pairedcritratio{}\,[\pairedcritratiolo,\pairedcritratiohi]$.

\begin{table*}[t]
\centering
\caption{Sample and accuracy at the reliability criterion \crit{}:
$N_{\rm ch}(\mathrm{S/N}>5)>50$, for the \zcut{} sample across all nine fields
($N_{\rm comp}=\nexternal{}$). \label{tab:retention}}
\small
\begin{tabular*}{\textwidth}{@{\extracolsep{\fill}}lccccc@{}}
\hline\hline
\colhead{Configuration} & \colhead{Median $S/N$} & \colhead{\ncrit} &
\colhead{\fcrit{} [\%]} & \colhead{\signmad} & \colhead{\etaout{} [\%]} \\
\hline
\linearest{} (\zcut) & $2.27$ & $2252$ & $28.9$ & $0.00823$ & $1.78$ \\
\efl{} (\zcut) & $2.27$ & $2253$ & $29.0$ & $0.00795$ & $1.55$ \\
\efl{} (full) & $1.06$ & $1357$ & $17.4$ & $0.00776$ & $1.40$ \\
\eflp{} (full) & $2.14$ & $2182$ & $28.0$ & $0.00906$ & $1.92$ \\
\hline
\end{tabular*}
\par
\begin{minipage}{\textwidth}
\tablecomments{Median S/N summarizes the per-source channel medians
before applying \crit{}. Redshift statistics use the
calibrated redshifts of each configuration on its own sample satisfying \crit{}.}
\end{minipage}
\end{table*}

The wider redshift scatter at full depth also persists at fixed magnitude,
though the catastrophic fractions agree within binomial uncertainties
(bottom right panels of Figure~\ref{fig:retention}). On sources calibrated at both depths,
the paired \signmad{} ratio is \pairedbinlo{}--\pairedbinhi{} in every
bin brighter than $\mz=\pairedbinmax{}$, and
$\pairedcalratio{}\,[\pairedcalratiolo,\pairedcalratiohi]$ over all
\npairedcal{} common sources. The calibration chain of each configuration
(Appendix~\ref{app:fields}) is included in this comparison. Although the floor does
not recover the performance of the truncated fit, it still improves on the
unregularized full-depth fit, which has larger scatter and catastrophic
fractions across the magnitude range.

However, if the reported uncertainties grew but the fluxes stayed
equally accurate, a significance requirement would
discard good measurements. We test this possibility on the simulated
scene of the depth experiment (Section~\ref{sec:sims}). For the
\nsimretcommon{} sources fitted at both depths, we replace the reported
uncertainties with their rms errors against the simulation truth, including
bias, measured over \nrealiz{} noise realizations. We then repeat the S/N
selection above \simretthreshold{}. With these errors, the \zcut{} fit yields \simretactual{} times as many
sources as the full-depth fit, compared with \simretreported{} using the reported uncertainties. Most of the sample
loss therefore reflects poorer flux measurements, although this
single-band A2537 test does not explain why the median S/N falls by a factor of
\medsnrdrop{} in the observational data but only by \simsnrdrop{} in the
simulated scene.

At the truncated depth, the floor changes little in the bulk of the
sample but removes a few catastrophic outliers. The \efl{} and
\linearest{} fits select nearly identical samples at \crit{}; on
\nlinearcommon{} common calibrated sources, the difference in
\signmad{} (\linearest{} minus \efl{}) is
$\lineardnmad{}\,\lineardnmadci{}$. Of these sources, \lineardiscord{} are
catastrophic outliers in the \linearest{} fit but not in the \efl{} fit;
no source is an outlier only in the \efl{} fit (exact
McNemar $p=\linearmcnemar$). Thus, at \zcut{}, the eigenvalue floor modestly
reduces the number of catastrophic outliers, whereas fitting the full
catalog still leads to a large loss of sources.

\subsection{External Priors: The Sample Back, Accuracy Lost}\label{sec:prior}

External flux priors constrain the contribution of faint neighbors and
can therefore reduce the neighbor light assigned to protected
targets \citep[e.g.,][]{pearson2017deblending}. We test whether this
constraint improves the target fluxes and recovers the sample lost at full catalog
depth, and assess the accompanying cost in accuracy.
The full-catalog \eflp{} configuration yields \ntierprior{} sources that
satisfy \crit{}, close to the \ntier{} of the baseline (Table~\ref{tab:retention}). Constraining the faint neighbors reduces
their freedom to exchange flux with the protected targets and narrows
the reported uncertainties of the targets, although the targets themselves
receive no flux prior. This narrowing comes from the priors rather than
from the \spherex{} images; if the expected fluxes are biased or the prior widths
are too narrow, the targets gain S/N without gaining accuracy. We therefore test whether
the prior configuration yields target fluxes as accurate as those of the baseline,
using the photometric redshifts of the sources calibrated in both
configurations and the broadband comparison of matched protected targets.

\begin{figure*}
\centering
\includegraphics[width=\textwidth]{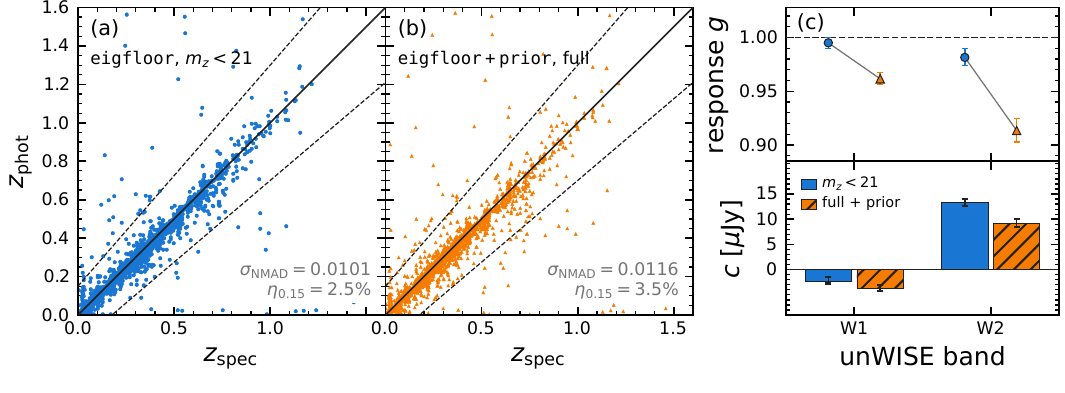}
\caption{Photometric redshift and broadband flux comparisons for the
\zcut{} \efl{} and full-catalog \eflp{} configurations.
(a, b) Photometric vs. spectroscopic redshifts for the same
\npriorpairnine{} sources, respectively, with separate calibration chains;
solid lines show equality, dashed lines $|\dz|=0.15$.
(c) Broadband response $g$ (upper) and offset $c$ (lower) relative to
unWISE W1 and W2 for matched protected targets, colored as in (a, b),
with 16th to 84th percentile bootstrap intervals.
The offset compares measurements limited by their respective catalog depths.
\label{fig:prioracc}}
\end{figure*}

The recovered sample comes with wider photometric-redshift scatter and
more catastrophic outliers (left and middle panels of Figure~\ref{fig:prioracc}). On the
\npriorpairnine{} sources calibrated in both configurations, the prior
increases \signmad{} by
$\priordnmadadopted{}\,\priordnmadadoptedci{}$ and \etaout{} by
$\priordcatadopted{}\,\priordcatadoptedci{}$ percentage points. There are \priorcatonly{} sources that are catastrophic only
with the prior, against \baselinecatonly{} that are catastrophic only in the
baseline.
The scatter difference is positive in each of the \npriorfieldsnonsparse{}
fields with at least \nsparsemin{} common sources; it also stays positive
when each field is given equal weight and when any single field is left out. Because the two configurations are compared on the same
sources, this accuracy cost reflects their photometry and calibration
(Appendix~\ref{app:fields}) rather than a difference in sample
composition.

In the broadband comparison (Equation~\ref{eq:broadband}), the prior
reduces the W2 offset $c$ but lowers the response $g$ in both WISE
bands (right panel of Figure~\ref{fig:prioracc}). The W2 offset
falls from \pedblind{}\,\ujy{} in the baseline to \pedprior{}\,\ujy{} with the
prior, by about a third; the decrease is significant in \wtwonresolved{} of
the \wtwonfields{} fields and when each field is given equal weight. Because
the expected fluxes in the prior use the W2 photometry, this comparison is
partly circular; excluding W2 from the prior keeps most of the decrease,
lowering the offset to \pedwithheld{}\,\ujy{}. The response $g$ is close to
unity in the baseline, \gwoneblind{} in W1 and \gwtwoblind{} in W2, and falls
to \gwoneprior{} and \gwtwoprior{} with the prior.
Appendix~\ref{app:diagnostics} details the fit without W2 in the prior and
shows how these changes vary with target brightness and band.

Varying the fractional prior width does not remove the accuracy cost.
The prior width sets how tightly the fit constrains the faint-neighbor fluxes.
In the COSMOS field, we test fractional widths from \priorscanmin{} to
\priorscanmax{}, keeping the minimum width at \priorsigmamin{}.
The broadband response changes little below the adopted
\priorsigmafrac{}, because this minimum prevents a tighter prior.
Separate redshift comparisons at
\priorsigmafrac{}, \priorscanmid{}, and \priorscanmax{} show no
significant gain in accuracy.

Better expected fluxes also leave the target fluxes nearly unchanged.
The interpolation used to set the expected neighbor fluxes can miss
spectral features that lie between the catalog bands. On \zcut{} sources with high-S/N
\spherex{} spectra, replacing interpolation with template fitting reduces
the discrepancy between predicted and measured fluxes at
\tmplwavelo{}--\tmplwavehi{}\,\um{} by a factor of \tmplbumpfactor{}.
We then set the expected flux of each faint neighbor from a template fit
to the catalog photometry that the interpolation uses and repeat the A1361
photometry with these priors. Over the same wavelength range, the
protected targets in the faintest quarter by $z$-band magnitude
brighten by a median of about \tmplfaintbumpfrac{} relative to the fit with
interpolated expected fluxes. The test is limited to A1361, and the
photometric redshifts with these priors remain untested.

\section{Discussion}\label{sec:discussion}

Here we compare our results with the two \spherex{} simulation studies
introduced in Section~\ref{sec:intro} and point to analogous depth tests for
other surveys (Section~\ref{sec:limit}). We then discuss the limits of this
measurement (Section~\ref{sec:limits}) and approaches beyond the two that we
test (Section~\ref{sec:beyond}).

\subsection{Catalog Depth and Photometric-Redshift Accuracy}\label{sec:limit}

\citet{huai2026confusion} studied how selection depth affects the number of
galaxies available for cosmology, finding that blending can make the
effective number density saturate as the catalog deepens. Here we focus on
performance differences among configurations, comparing each pair of fits
on the targets they share.
This approach keeps changes in sample composition from obscuring those
differences; the counts at \crit{} separately measure the sample available
from each fit. Our observational measurements thus complement their forecasts of the
usable cosmology sample. The same blending mechanism drives catastrophic
outliers in the galaxy-cluster simulations of \citet{bahk2026clusters}, which were
designed to reproduce realistic \spherex{} observations.
At their criterion, our baseline gives $\signmad=\bahktierhere$ compared
with their \bahktierpub{}.
Although the samples and calibration procedures differ, obtaining
this lower scatter before the full \surveyyears{}-year survey is complete is encouraging
for \spherex{} photometric-redshift accuracy.

The limit on how many source fluxes the images constrain
(Section~\ref{sec:identifiability}) arises whenever a deeper catalog adds source
profiles that the lower-resolution pixels cannot distinguish. It therefore
motivates analogous depth tests for \euclid{} photometry with high-resolution
source priors \citep{romelli2025euclidmer} and for extracting overlapping
\romantel{} slitless spectra \citep{gabrielpillai2026espresso}.
The relevant boundary depends on source density, angular resolution, and
exposure depth. The fitted catalog should be matched to what those data can
constrain, with the choice checked against source truth or an end-to-end
science metric.

\subsection{Limits of This Measurement}\label{sec:limits}

The scope of the measured trade-off is limited by the sample,
calibration, and simulation design. First, the comparison sample is
brighter than the full survey population and is concentrated in the COSMOS
field
(Section~\ref{sec:cohort}), but the Coma field dominates the paired redshift samples
(Appendix~\ref{app:sample-audit}). Nevertheless, every paired scatter difference keeps
its sign when the fields are weighted equally or when any one field is left
out.
The coverage requirement also
restricts the footprint to regions observed at least as often as the
field core. Second, because each configuration uses its own calibration pool and
cross-validation folds (Appendix~\ref{app:photoz-calibration}), paired redshift differences include the
calibration response. These differences are insensitive at least to
the channel responses, changing little when every calibration step is
derived again with laboratory-measured responses
(Appendix~\ref{app:lab-response}). Third, the depth simulation varies only the noise
in one scene and one band under a shared rendering and fitting model;
it does not sample other source populations or profile errors.
However, the cost of fitting the full catalog does not rest on this
simulation, because we measure it directly in the observational data, which
include the real source populations and profiles
(Section~\ref{sec:retention}).

Morphology illustrates another limitation of using redshifts to assess
photometry. \citet{bahk2026clusters} found that treating simulated
galaxies as point sources underestimated their fluxes even for
half-light radii below \bahkpsfradius{}, typically by \bahkpsfbias{} and by as much
as \bahkpsfbiasmax{}. In our test on observational data, replacing galaxy profiles
with point sources also lowers the fluxes of resolved galaxies, and
the decrease grows with size, reaching \allpsfextloss{} for half-light radii
above \allpsfextradius{}. However, the change in the bulk scatter of
calibrated redshifts is consistent with zero. Three factors can mask this flux change in the calibrated
redshifts. Compact sources dominate the sample; the template normalization
can absorb an overall flux rescaling; the redshift calibration can absorb
some morphology-dependent shifts. Agreement
in calibrated redshifts therefore does not establish the accuracy of the
fluxes or the assumed profiles.

The full-survey processing times in Section~\ref{sec:engine} are
estimates based on A2537, but the actual times depend on the source
density and the available computing resources; the density of A2537, which
exceeds the survey average, makes the estimates conservative. The
estimates also exclude downloading the images and saving the output
catalogs. Because these steps run on the network and the CPU, they can
proceed during the GPU fits (Section~\ref{sec:model}) and are expected to
add little to the processing time. Even allowing for these caveats on the
timing, the speed gain of the GPU is large enough to make repeated fits
practical for both full-survey reprocessing and photometry tailored to
individual science goals.

\subsection{Beyond the Two Approaches}\label{sec:beyond}

Three directions could extend the fit past the truncation boundary or
test where that boundary lies. First, the fit could use fewer free
parameters by grouping unresolved neighbors into shared components or by
tying the channel fluxes to a small set of spectral parameters; the accuracy
of either would depend on its modeling assumptions. Second, priors built
from deeper external photometry in more bands could constrain faint
neighbors with information absent from our optical and WISE inputs.
However, deep multiband
data of this kind remain unavailable over much of the all-sky footprint of \spherex{}.
Third, the \spherex{} deep fields would test how the truncation boundary
shifts with exposure depth, particularly at shallow catalog cuts where
\citet{huai2026confusion} forecast that spectral confusion dominates over
blending in degrading redshifts.

\section{Summary}\label{sec:summary}

We have developed \tractorjax{}, a GPU reimplementation of
the catalog-forced mode of \thetractor{}, and used it to examine the
computational cost and statistical limits of \spherex{} forced
photometry. Using simulations and observational data, we assess how fitted-catalog depth, regularization,
and external flux priors affect flux recovery, the number of reliably
measured sources, and photometric-redshift accuracy.

\begin{enumerate}
\item
\tractorjax{} reports uncertainties verified against simulation truth and
agrees with \thetractor{} to a median flux ratio within \parityaltmed{} of unity.
Its full-depth service time is \eigservicems{}\,ms per cutout on one NVIDIA H100 GPU,
against \cpucoresec{}\,s per CPU core. At these rates, a mission pass at the benchmark
density would take about \missioncarddays{} GPU-days versus
\missioncoreyr{} core-years, excluding image downloads and catalog writes.

\item
In the truth-controlled A2537 simulation, opening the fitted catalog from
\zcut{} to full \ls{} depth mainly adds sources that are too faint to constrain
and widens the scatter of the fractional flux residuals of the common
sources by a factor of \popscatterratio{}.
Without regularization, the factor is \popscatterlinratio{}. Because truncated
depths within $\mz=\plateaurange$ span scatter ratios of only
\scatterrunglo{}--\scatterrunghi{} relative to the baseline, the precise
boundary matters less than fitting beyond the individually
measurable population.

\item
On the observational comparison sample, full-depth regularization alone lowers
the channel S/N through larger flux uncertainties, leaving \ntierfull{} sources that
satisfy \crit{} (Equation~\ref{eq:criterion}) against \ntier{} for the truncated fit.
The truncated fit supplies \amratiolo{}--\amratiohi{} times more sources
at the same required photometric-redshift scatter; on sources meeting \crit{}
at both depths, \signmad{} of the full-depth fit is \pairedcritratio{} times larger.

\item
External flux priors recover \ntierprior{} sources at \crit{}, but on
the common calibrated sources, they raise \signmad{} by \priordnmadadopted{}
and the catastrophic fraction by \priordcatadopted{} percentage points
relative to the baseline.
They reduce the W2 offset relative to unWISE from
\pedblind{} to \pedprior{}\,\ujy{} but lower the response in both
WISE bands. The tested prior configuration therefore recovers most of the
sample at a cost in target accuracy.
\end{enumerate}

Fitted-catalog depth should therefore be assessed in terms of the
sample size and accuracy it delivers for the intended science.
\tractorjax{} reduces the cost of repeatedly testing that choice,
from reprocessing the full survey to extracting photometry for
individual research projects.

\restartappendixnumbering
\begin{acknowledgments}
We thank Yujin Yang, Jeonghyun Pyo, and Woong-Seob Jeong for their
advice on the use of \spherex{} data products and for their support
throughout this project.
The work of HB was supported by Basic Science Research Program through the
National Research Foundation of Korea (NRF) funded by the Ministry of Education
(RS-2025-25403440). HSH acknowledges support from the National Research
Foundation of Korea (NRF) funded by the Korea government (MSIT; RS-2026-25482692)
and the Global-LAMP Program funded by the Ministry of Education (RS-2023-00301976).
This publication makes use of data products from the Spectro-Photometer
for the History of the Universe, Epoch of Reionization and Ices Explorer
(\spherex{}), which is a joint project of the Jet Propulsion Laboratory and
the California Institute of Technology, and is funded by the National
Aeronautics and Space Administration.
\end{acknowledgments}

\facilities{\spherex{}}

\software{\tractorjax{} \citep[this work; v0.3.0,][]{bahk2026tractorjax},
\thetractor{} \citep{lang2016tractor},
\eazy{} \citep{brammer2008eazy},
\jax{} \citep{jax2018github},
\galsim{} \citep{rowe2015galsim},
Astropy \citep{astropy2022},
NumPy \citep{harris2020numpy},
SciPy \citep{virtanen2020scipy},
Matplotlib \citep{hunter2007matplotlib},
photutils \citep{bradley2024photutils}}

\appendix

\section{Photometric-Redshift Calibration and Sample Definitions}\label{app:fields}

The redshift comparisons depend on the calibration and sample selection
of each configuration. Here we describe the calibration steps and
whether each is derived without the evaluation sources, the spectroscopic
sources on which the redshift accuracy is measured
(Appendix~\ref{app:photoz-calibration}). We then explain the
differences in sample size (Appendix~\ref{app:sample-audit}) and separate the
contribution of the template-error amplitude from the redshift cost of the
prior (Appendix~\ref{app:tef-amplitude}). Finally, we repeat the redshift
analysis with laboratory-measured channel responses
(Appendix~\ref{app:lab-response}).

\subsection{Calibration Chain}\label{app:photoz-calibration}

Photometric-redshift estimates require calibration because systematic
flux offsets and mismatches between the observed spectra and the
templates can bias the inferred redshifts
\citep{salvato2019photoz}. We calibrate each photometry configuration
separately, where a configuration specifies both the estimator
(\efl{}, \linearest{}, \eflp{}, or \lassoest{}) and the fitted catalog
depth (\zcut{} or full depth).
We first adjust the channel zeropoints against
template predictions at fixed \zspec{}, following the spectroscopic
calibration approach of \citet{ilbert2006photoz}, and allow for residual
spectral mismatch through template-error terms \citep{brammer2008eazy}.
Table~\ref{tab:appcalibration} lists each stage, the data it uses, and
whether it excludes the evaluation sources.

\providecommand{\rg}{\raggedright\let\\\tabularnewline}
\begin{deluxetable*}{p{0.21\textwidth}p{0.36\textwidth}p{0.34\textwidth}}
\tabletypesize{\footnotesize}
\tablecaption{Photometric-redshift calibration stages and the samples used to derive them.
\label{tab:appcalibration}}
\tablehead{\colhead{Stage} & \colhead{Calibration Data} &
\colhead{Evaluation Sources Excluded?}}
\startdata
\rg Channel zeropoints & \rg The protected spectroscopic pool for that
estimator and catalog depth, using template residuals at fixed \zspec{} & \rg No: the final
channel scales use the complete pool; cross-validation is a separate check. \\
\rg S/N fitting weight & \rg Truth-controlled simulations; a fixed error
multiplier shared by all configurations & \rg Independent of the
spectroscopic folds; no residual curve from the observational data is fitted. \\
\rg Template-error curves and amplitude & \rg Rest- and observed-frame
residuals on the complete protected pool; amplitude from the rule of Appendix~\ref{app:tef-amplitude} & \rg No: curve estimation and amplitude selection use
the complete pool for that estimator and catalog depth. \\
\rg Point warp, $z_{\rm raw}\to\zphot$ & \rg Photometric and spectroscopic
redshifts in the other four folds & \rg Yes: applied only to the fold
left out of the correction. \\
\enddata
\tablecomments{Evaluation sources are the spectroscopic sources on which the
redshift accuracy is measured. Channel scales multiply both fluxes and errors; the S/N
multiplier changes only the template-fitting weights.
Folds are assigned separately for each photometry estimator and catalog
depth and are not shared across matched sources.}
\end{deluxetable*}

The template-fitting weights also include a fixed, simulation-derived
multiplier of the flux error, which at channel S/N $s$ is
\begin{equation}
 r(s)=\snweightfunction,
 \label{eq:snweight}
\end{equation}
where clipping confines the result to the listed bounds.
\citet{bahk2026clusters} found that channel fluxes averaged over visits
scatter about \bahkpullwidth{} times more than their formal errors, mainly
because the visits sample each channel at slightly different wavelengths;
the multiplier compensates for this underestimate.

The \emph{point warp} corrects the typical redshift bias as a function
of the raw template-fit redshift $z_{\rm raw}$.
After excluding catastrophic outliers, we compute median normalized
redshift residuals in equally populated bins of $z_{\rm raw}$ and connect
these medians with shape-preserving cubic interpolation.
The resulting curve $d(z)$ gives
$\zphot=z_{\rm raw}-d(z_{\rm raw})(1+z_{\rm raw})$, with the endpoint
corrections held constant outside the fitted range.

To exclude each evaluation source from the warp applied to it, we
randomly divide the spectroscopic sample of each photometry configuration
into five folds, derive the warp from four of them, and apply it to the
remaining fold in turn.
For the full calibration sample of the baseline (\nspecz{} sources), the warp
reduces \signmad{} in all five evaluation folds, with corrected median
\dz{} ranging from \warpfoldbiaslo{} to \warpfoldbiashi{}.
The channel zeropoints, the template-error curves, and the
template-error amplitude (Appendix~\ref{app:tef-amplitude}) are instead
fitted to the complete spectroscopic pool for each estimator and catalog
depth.
Because folds are assigned separately for each estimator and catalog depth,
a source matched between two configurations is corrected by warps from
different folds. A paired difference therefore reflects the calibration of
each configuration as well as its photometry (Section~\ref{sec:photoz}).

\subsection{Sample Definitions}\label{app:sample-audit}

Table~\ref{tab:appaudit} distinguishes the full
calibrated sample of each configuration from the sources used to compare configurations.
The \emph{comparison sample} is the \zcut{} population defined in
Section~\ref{sec:cohort}, with $N_{\rm comp}$ sources.
$N_{\rm native}$ counts the calibration pool of each configuration
(Section~\ref{sec:photoz}): the spectroscopic sources brighter than \zbound{}
whose median channel \snpresel{}, all of which receive a calibrated
photometric redshift. Unlike the comparison sample, the pool has no coverage
requirement and therefore also contains sources observed less often than
the field core. $N_{\rm valid}$ counts the sources in both, that is, the
members of the comparison sample that enter the pool. These sources form the
\emph{valid sample} of the configuration.

\providecommand{\rg}{\raggedright\let\\\tabularnewline}
\begin{deluxetable*}{p{0.16\textwidth}ccccccc}
\tabletypesize{\footnotesize}
\tablecaption{Sample sizes and paired redshift differences for the five
main-text configurations. \label{tab:appaudit}}
\tablehead{\colhead{Configuration} & \colhead{$N_{\rm native}$} &
\colhead{$N_{\rm valid}$} & \colhead{$N_{\rm common}$} &
\colhead{$\Delta\signmad$} & \colhead{$\Delta\signmad$} &
\colhead{$\Delta\etaout$ [pp]} & \colhead{$a/b$} \\
\colhead{} & \colhead{} & \colhead{} & \colhead{} &
\colhead{Pooled [95\%]} & \colhead{\shortstack{Equal-field\\(Pooled LOFO)}} &
\colhead{Pooled [95\%]} & \colhead{}}
\startdata
\rg \linearest{} (\zcut) & $4416$ & $3203$ & $3203$ & \begin{tabular}[t]{@{}c@{}}$+0.00009$\\$[-0.00009,+0.00031]$\end{tabular} & \begin{tabular}[t]{@{}c@{}}$+0.00042$\\$(+0.00004,+0.00050)$\end{tabular} & \begin{tabular}[t]{@{}c@{}}$+0.22$\\$[+0.06,+0.41]$\end{tabular} & $7/0$ \\
\rg \efl{} (\zcut) & $4419$ & $3205$ & $3205$ & \nodata & \nodata & \nodata & \nodata \\
\rg \linearest{} (full) & $3462$ & $2617$ & $1985$ & \begin{tabular}[t]{@{}c@{}}$+0.00423$\\$[+0.00341,+0.00517]$\end{tabular} & \begin{tabular}[t]{@{}c@{}}$+0.03919$\\$(+0.00297,+0.02421)$\end{tabular} & \begin{tabular}[t]{@{}c@{}}$+6.75$\\$[+5.84,+7.66]$\end{tabular} & $138/4$ \\
\rg \efl{} (full) & $2840$ & $2004$ & $1992$ & \begin{tabular}[t]{@{}c@{}}$+0.00251$\\$[+0.00187,+0.00324]$\end{tabular} & \begin{tabular}[t]{@{}c@{}}$+0.00341$\\$(+0.00227,+0.00541)$\end{tabular} & \begin{tabular}[t]{@{}c@{}}$+0.25$\\$[-0.10,+0.65]$\end{tabular} & $10/5$ \\
\rg \eflp{} & $4277$ & $3070$ & $3034$ & \begin{tabular}[t]{@{}c@{}}$+0.00151$\\$[+0.00098,+0.00215]$\end{tabular} & \begin{tabular}[t]{@{}c@{}}$+0.00369$\\$(+0.00141,+0.00295)$\end{tabular} & \begin{tabular}[t]{@{}c@{}}$+0.92$\\$[+0.46,+1.38]$\end{tabular} & $40/12$ \\
\enddata
\tablecomments{$N_{\rm native}$ counts calibrated redshifts in the
calibration pool of each configuration; $N_{\rm valid}$ counts those within the
\zcut{} comparison sample ($N_{\rm comp}=7781$) that pass the
median-channel \snpresel{} selection.
$N_{\rm common}$ counts the sources included in $N_{\rm valid}$ for both that
configuration and the baseline, before applying \crit{}.
Contrasts subtract the \zcut{} \efl{} baseline; square brackets give 95\%
within-field paired-bootstrap intervals, and $a$ and $b$ count catastrophic
outliers unique to the listed configuration and baseline, respectively.
Equal-field values average contrasts over fields with at least \nsparsemin{}
common sources; parentheses give the pooled leave-one-field-out (LOFO)
range, recalculated after omitting each field in turn without excluding
sparse fields.}
\end{deluxetable*}

Each configuration is compared with the \zcut{} \efl{} baseline on
the sources in the valid samples of both configurations.
This \emph{paired sample} contains $N_{\rm common}$ sources. All counts in
Tables~\ref{tab:appaudit} and~\ref{tab:appfieldcounts} precede the
reliability criterion \crit{}.
Table~\ref{tab:appfieldcounts} gives the counts for each field, including
the sparsest galaxy-cluster samples. Because the Coma field
dominates every paired sample, Table~\ref{tab:appaudit} also gives each
scatter difference with equal field weights and with one field left out.
Every difference keeps its sign in both tests and is larger with equal
weights than in the pooled sample; omitting Coma gives the largest pooled
value.

\begin{deluxetable*}{lccccccc}
\tablecolumns{8}
\tabletypesize{\footnotesize}
\tablecaption{Sample sizes by field for the depth and prior comparisons.
\label{tab:appfieldcounts}}
\tablehead{\colhead{Field} & \colhead{$N_{\rm comp}$} &
\multicolumn{3}{c}{$N_{\rm valid}$} & \colhead{} &
\multicolumn{2}{c}{$N_{\rm common}$} \\
\cline{3-5}\cline{7-8}
\colhead{} & \colhead{} & \colhead{Baseline} &
\colhead{\efl{} (full)} & \colhead{\eflp} & \colhead{} &
\colhead{\efl{} (full)} & \colhead{\eflp}}
\startdata
COSMOS & $4337$ & $700$ & $199$ & $641$ &  & $199$ & $636$ \\
Coma & $2899$ & $2143$ & $1531$ & $2078$ &  & $1519$ & $2053$ \\
A2055 & $230$ & $186$ & $154$ & $184$ &  & $154$ & $180$ \\
A1361 & $48$ & $42$ & $36$ & $40$ &  & $36$ & $40$ \\
A2187 & $50$ & $45$ & $37$ & $45$ &  & $37$ & $44$ \\
A2537 & $170$ & $70$ & $40$ & $67$ &  & $40$ & $66$ \\
SPT-CL J2145$-$5644 & $21$ & $7$ & $2$ & $6$ &  & $2$ & $6$ \\
SpARCS J1613$+$5649 & $19$ & $10$ & $5$ & $8$ &  & $5$ & $8$ \\
SPT-CL J0546$-$5345 & $7$ & $2$ & $0$ & $1$ &  & $0$ & $1$ \\
\enddata
\tablecomments{$N_{\rm comp}$ counts sources in the \zcut{} comparison sample
for each field.
Each common sample is the intersection of the valid samples of that
configuration and the baseline. All counts precede the reliability
criterion \crit{}.}
\end{deluxetable*}

\begin{figure*}
\centering
\includegraphics[width=\textwidth]{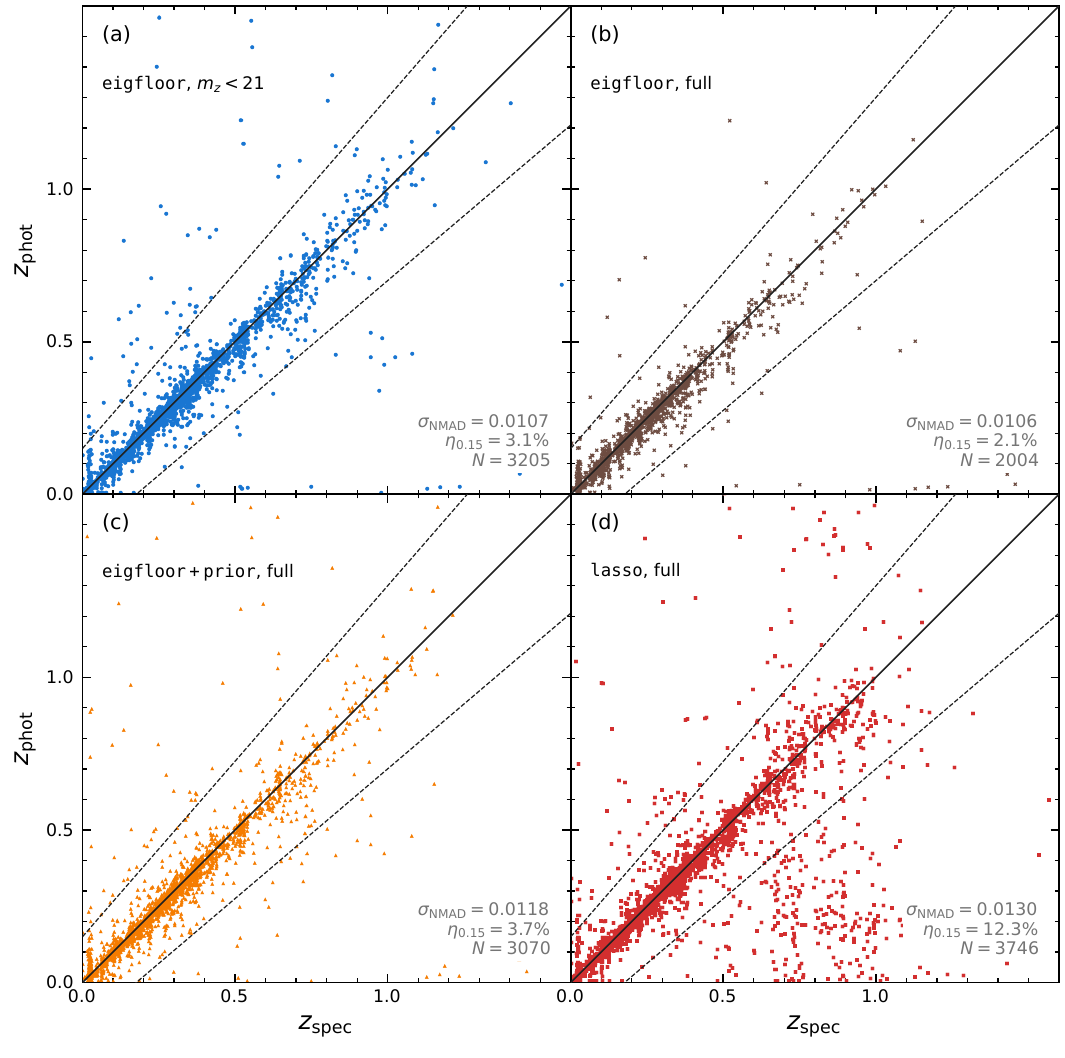}
\caption{Photometric vs. spectroscopic redshifts for the
valid sample of each configuration.
Solid line: identity; dashed: $|\dz|=0.15$.
Annotations give the \signmad{}, \etaout{}, and size
$N=N_{\rm valid}$ (Appendix~\ref{app:sample-audit}) of each sample, including
sources outside the frame.
Comparisons on common sources are shown in
the left and middle panels of Figure~\ref{fig:prioracc}.
\label{fig:zzarms}}
\end{figure*}

Figure~\ref{fig:zzarms} shows the calibrated redshifts of each configuration
for its valid sample.
Differences between panels therefore reflect both the redshift estimates and
which sources enter each valid sample, whereas comparisons on the same sources use
the paired samples defined above.

\subsection{Template-Error Amplitude}\label{app:tef-amplitude}

Every configuration selects the template-error function amplitude
(\atefsym{}) with the same rule. For each \atefsym{} on a grid from zero to
\atefgridmax{}, we compute the difference in \signmad{} from the lowest value
on the grid, on the same sources, with a \atefplateaulevel{} bootstrap
interval. The rule selects the smallest \atefsym{} whose difference is
consistent with zero: the smallest value on the \signmad{} plateau.
This choice avoids larger values on the plateau, which increase the
bright-source redshift scatter of the baseline without a measurable
improvement in the full-sample scatter.
The template-error term has greater influence at high S/N, where the
photometric errors are small.
With this rule, every configuration selects $\atefsym=\atefopt{}$.
Table~\ref{tab:prior} separates the effect of the amplitude from the
redshift cost of the prior. If only the prior configuration used
$\atefsym=\atefshipped{}$, with its warp recalibrated on the same sources, its
\signmad{} difference from the baseline would rise from
\priordnmadadopted{} to \priordnmad{}, and the amplitude would account for
\priorsharedfrac{} of this difference.

\providecommand{\rg}{\raggedright\let\\\tabularnewline}
\begin{deluxetable*}{lccc}
\tabletypesize{\footnotesize}
\tablecaption{Paired redshift differences for the prior configuration
at two template-error amplitudes. \label{tab:prior}}
\tablehead{\colhead{Contrast at the listed $\atefsym$} &
\colhead{$\Delta\signmad$ [95\%]} & \colhead{$\Delta\etaout$ [pp, 95\%]} &
\colhead{$a/b$}}
\startdata
\rg \eflp{} at $\atefoptprior{}$ $-$ baseline & \begin{tabular}[t]{@{}c@{}}$+0.00151$\\ $[+0.00098,+0.00215]$\end{tabular} & \begin{tabular}[t]{@{}c@{}}$+0.92$\\ $[+0.46,+1.38]$\end{tabular} & $40/12$ \\
\rg \eflp{} at $\atefshipped{}$ $-$ baseline & \begin{tabular}[t]{@{}c@{}}$+0.00314$\\ $[+0.00249,+0.00382]$\end{tabular} & \begin{tabular}[t]{@{}c@{}}$+1.55$\\ $[+0.96,+2.14]$\end{tabular} & $66/19$ \\
\rg \eflp{} at $\atefshipped{}$ $-$ at $\atefoptprior{}$ & \begin{tabular}[t]{@{}c@{}}$+0.00163$\\ $[+0.00104,+0.00216]$\end{tabular} & \begin{tabular}[t]{@{}c@{}}$+0.63$\\ $[+0.13,+1.12]$\end{tabular} & $41/22$ \\
\enddata
\tablecomments{All rows use the same \npriorpairnine{} common \zcut{}
sources, with the baseline fixed at $\atefsym=\atefopt{}$ and only the
amplitude of the prior configuration varied, followed by its warp recalibration.
The second row equals the sum of the first and third within every paired
bootstrap draw, so the third accounts for 52\% of the
scatter difference in the second row.
Square brackets are 95\% field-stratified paired-bootstrap intervals.
$a$ and $b$ count sources that are catastrophic outliers only in the fit
before and only in the fit after the minus sign, respectively.
The first row is the adopted contrast in Table~\ref{tab:appaudit}.}
\end{deluxetable*}

\subsection{Laboratory Channel Responses}\label{app:lab-response}

Replacing the Gaussian channel responses of Section~\ref{sec:photoz} with
laboratory-measured ones leaves the main redshift comparisons unchanged
(Table~\ref{tab:appresponse}). The measured responses
\citep{hui2026spectral} were
released\footnote{\url{https://irsa.ipac.caltech.edu/data/SPHEREx/docs/spherex_spectral_calibrations.html}}
after this analysis was completed. They have flatter tops and steeper
edges than a Gaussian of equal width, but their resolving powers agree
with the adopted band-plan values to within \labrtol{} in all but one
band-edge channel. The release gives two responses for each channel:
that of its central pixel and the mean over all of its pixels. We repeat the
redshift analysis with the all-pixel means, because our channel fluxes
average visits that fall on different pixels of each channel. For the
baseline, the central-pixel responses give nearly identical redshifts. Each response is shifted to the pivot wavelength of its Gaussian
counterpart, so that only the shape differs. Every calibration step
is derived again under the same rules, whereas the spectra, and hence the samples
satisfying \crit{}, are identical by construction.

The two response sets give the same redshifts to well within the scatter.
Through the template library they predict channel fluxes that differ by a
median of \labfluxmed{}. The NMAD of
$(z_{\rm lab}-z_{\rm Gaussian})/(1+\zspec)$ is
\labdznmadlo{}--\labdznmadhi{}, about a tenth of the photometric-redshift
scatter. At \crit{}, \signmad{} changes by at most \labdsigmax{};
the 95\% interval of every difference contains zero. Every
configuration again selects $\atefsym=\atefopt{}$. With the laboratory
responses, the prior increases \signmad{} by
$\labpriordnmad{}\,\labpriordnmadci{}$, against \priordnmadadopted{} with
the Gaussian responses. On sources meeting \crit{} at both depths,
\signmad{} of the full-depth fit is larger by a factor of
$\labpairedcritratio{}\,[\labpairedcritratiolo,\labpairedcritratiohi]$,
against \pairedcritratio{}.

\providecommand{\rg}{\raggedright\let\\\tabularnewline}
\begin{deluxetable*}{p{0.14\textwidth}ccccccc}
\tabletypesize{\footnotesize}
\tablecaption{Redshift statistics with Gaussian and with laboratory-measured
channel responses. \label{tab:appresponse}}
\tablehead{\colhead{Configuration} &
\multicolumn{3}{c}{\signmad{} at \crit} &
\multicolumn{2}{c}{\etaout{} at \crit{} [\%]} &
\multicolumn{2}{c}{$\Delta\signmad$ against the baseline [95\%]} \\
\colhead{} & \colhead{Gaussian} & \colhead{Lab} &
\colhead{Difference [95\%]} & \colhead{Gaussian} & \colhead{Lab} &
\colhead{Gaussian} & \colhead{Lab}}
\startdata
\rg \linearest{} (\zcut) & $0.00823$ & $0.00804$ & \begin{tabular}[t]{@{}c@{}}$-0.00019$\\$[-0.00039,+0.00019]$\end{tabular} & $1.78$ & $1.69$ & \begin{tabular}[t]{@{}c@{}}$+0.00009$\\$[-0.00009,+0.00031]$\end{tabular} & \begin{tabular}[t]{@{}c@{}}$+0.00002$\\$[-0.00013,+0.00023]$\end{tabular} \\
\rg \efl{} (\zcut) & $0.00795$ & $0.00804$ & \begin{tabular}[t]{@{}c@{}}$+0.00009$\\$[-0.00017,+0.00031]$\end{tabular} & $1.55$ & $1.51$ & \nodata & \nodata \\
\rg \efl{} (full) & $0.00776$ & $0.00808$ & \begin{tabular}[t]{@{}c@{}}$+0.00031$\\$[-0.00011,+0.00051]$\end{tabular} & $1.40$ & $1.40$ & \begin{tabular}[t]{@{}c@{}}$+0.00251$\\$[+0.00187,+0.00324]$\end{tabular} & \begin{tabular}[t]{@{}c@{}}$+0.00278$\\$[+0.00209,+0.00342]$\end{tabular} \\
\rg \eflp{} (full) & $0.00906$ & $0.00898$ & \begin{tabular}[t]{@{}c@{}}$-0.00007$\\$[-0.00029,+0.00029]$\end{tabular} & $1.92$ & $1.92$ & \begin{tabular}[t]{@{}c@{}}$+0.00151$\\$[+0.00098,+0.00215]$\end{tabular} & \begin{tabular}[t]{@{}c@{}}$+0.00164$\\$[+0.00101,+0.00226]$\end{tabular} \\
\enddata
\tablecomments{\signmad{} and \etaout{} are evaluated on the
sample of each configuration satisfying \crit{}, as in Table~\ref{tab:retention};
that sample is identical for the two response sets. The difference is laboratory
minus Gaussian on the same sources. The last two columns repeat the pooled
paired contrast of Table~\ref{tab:appaudit} for each response set. Brackets
are 95\% within-field paired-bootstrap intervals.}
\end{deluxetable*}

\section{Broadband Tests of External Flux Priors}\label{app:diagnostics}

We test whether the reduction of the W2 offset depends on reusing W2 by
withholding that band from the prior.
On the matched protected targets, the pooled W2 offset falls from
\pedblind{}\,\ujy{} in the baseline to \pedprior{}\,\ujy{} with the standard
prior and to \pedwithheld{}\,\ujy{} with W2 withheld
(Table~\ref{tab:appprior}). The reduction therefore does not rely solely on
reusing W2; withholding W2 raises the offset by only
\withhelddc{}\,\withhelddcci{}\,\ujy{} relative to the standard prior
(paired-bootstrap \anchorpercentiles{} percentiles). Both prior fits keep a positive W2 offset and lower the response in
both WISE bands. The withheld fit is a diagnostic rather than an alternative
configuration; the \ls{} $z$ and W1 bands still enter both its prior and the
comparison.

\providecommand{\mch}[1]{\multicolumn{1}{c}{\rule[-2pt]{0pt}{13pt}#1}}
\providecommand{\mcu}[1]{\multicolumn{1}{c}{\rule[-5pt]{0pt}{10pt}#1}}
\begin{deluxetable*}{ccccccccccc}
\tablecolumns{11}
\tabletypesize{\footnotesize}
\tablecaption{Broadband flux comparisons for the baseline, standard prior,
and W2-withheld prior.
\label{tab:appprior}}
\tablehead{\colhead{Band} &
\multicolumn{2}{c}{\efl} & \colhead{} &
\multicolumn{3}{c}{\eflp} & \colhead{} &
\multicolumn{3}{c}{W2 withheld} \\
\cline{2-3}\cline{5-7}\cline{9-11}
\mcu{} & \mcu{$g$} & \mcu{$c$ [\ujy]} & \mcu{} & \mcu{$g$} &
\mcu{$c$ [\ujy]} & \mcu{$\Delta c$ [\ujy]} & \mcu{} & \mcu{$g$} &
\mcu{$c$ [\ujy]} & \mcu{$\Delta c$ [\ujy]}}
\startdata
W1 & $0.995$ & $-2.25$ & & $0.962$ & $-3.69$ &
$-1.44\;[-2.01,-0.98]$ & & $0.961$ & $-3.95$ &
$-1.71\;[-2.25,-1.22]$ \\
W2 & $0.982$ & $+13.37$ & & $0.914$ & $+9.18$ &
$-4.19\;[-4.83,-3.50]$ & & $0.889$ & $+9.72$ &
$-3.65\;[-4.39,-2.89]$ \\
\enddata
\tablecomments{Each band uses a random subsample of 8000 galaxies common
to all fits, selected from all nine fields with the magnitude, morphology,
and coverage requirements of Section~\ref{sec:cohort}.
This sample does not require the spectroscopic redshifts used to define
$N_{\rm comp}$;
sources must have finite fluxes in every fit and
$0<F_{\rm unWISE}<\anchorfluxmax{}\,\ujy$.
The Theil--Sen response $g$ and offset $c$ follow Equation~\ref{eq:broadband},
with $\Delta c$ calculated by subtracting the baseline $c$ from the value
for each prior fit. Brackets give \anchorpercentiles{} percentile intervals
from \anchorboot{} paired bootstrap resamples.}
\end{deluxetable*}

\begin{figure}[h]
\setlength{\hsize}{\textwidth}
\setlength{\linewidth}{\textwidth}
\centering
\includegraphics[width=\textwidth]{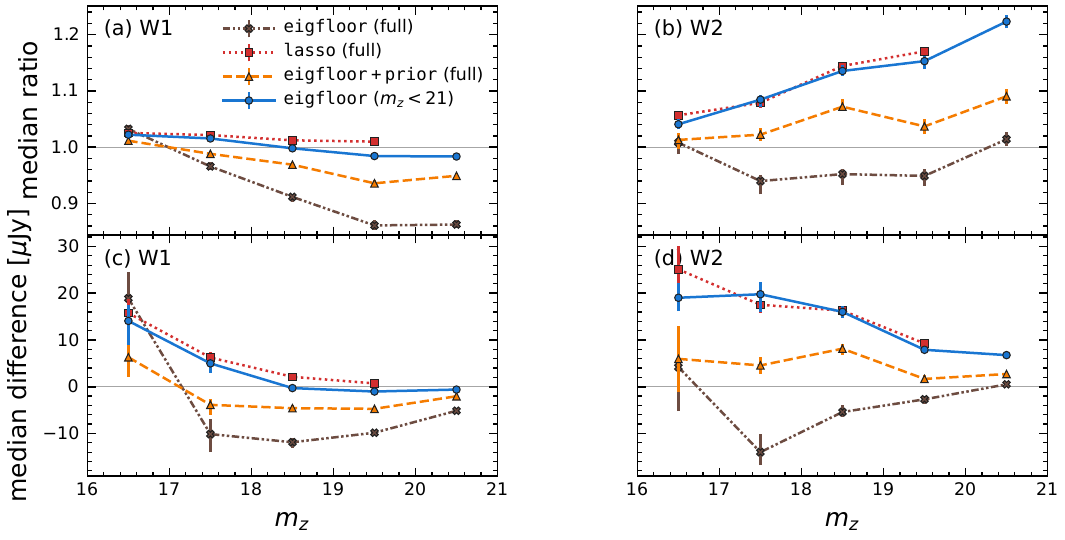}
\caption{Broadband flux comparisons between \spherex{} and unWISE in W1
and W2 as a function of target magnitude. All matched sources in each magnitude
bin are pooled across the nine fields, without the random subsampling used
in Table~\ref{tab:appprior}.
Panels (a, b) show median $F_{\mathrm{\spherex}}/F_{\rm unWISE}$ and
(c, d) median $F_{\mathrm{\spherex}}-F_{\rm unWISE}$ in one-magnitude
bins of \mz{}, with paired-bootstrap
\anchorpercentiles{} percentile intervals. The sparse fit is shown only
for \lassocut{}. These comparisons of observational data measure differences between
catalog-depth-limited measurements, not absolute flux errors.
\label{fig:anchormag}}
\end{figure}

Comparisons in magnitude bins show which sources are most affected
by the prior. The prior reduces the excess of the baseline W2 fluxes over
unWISE across target
magnitudes, with a larger decrease in the flux ratio toward fainter targets
and redder bands (Figure~\ref{fig:anchormag}). The regularized
full-depth fit also reduces this excess.
The sparse fit follows the baseline for \lassocut{} but zeroes
\lassozerofrac{}\% of W1/W2 targets in the next magnitude bin
(Appendix~\ref{app:lasso}).

\onecolumngrid
\newpage
\twocolumngrid

\section{Sparse Selection in Place of Truncation}\label{app:lasso}

Sparse selection keeps faint neighbors in the fitted catalog but adds an
$\ell_1$ penalty on their fluxes, which sets a flux to zero when the data do
not support it. The data therefore select which neighbors remain in the fit,
reducing the flux degeneracy without a fixed magnitude cut. We protect the
\lassocut{} targets from the penalty, so that only fainter sources can be
set to zero. We apply sparse selection at full catalog depth, where it
replaces truncation, and use a fit at \zcut{} as a same-depth control. We
then ask whether the full-depth sparse fit yields more sources with
calibrated redshifts than the baseline and whether those redshifts are
accurate.

We implement sparse selection with the least absolute shrinkage and
selection operator (LASSO; \citealt{tibshirani1996lasso,wang2014desphot}).
For the model in Equation~\ref{eq:forward},
the image residual is $\boldsymbol r=\boldsymbol d-A\boldsymbol f-B\boldsymbol b$.
The Gaussian noise model gives a negative log-likelihood
$\chi^2/2=\boldsymbol r^{\top}W\boldsymbol r/2$, up to an additive constant.
The LASSO minimizes
$\chi^2/2+\mathcal P_1$, with the penalty
\begin{equation}
 \mathcal P_1=\alpha\sum_j w_j\sqrt{G_{jj}}\,|f_j|,
 \label{eq:lasso-penalty}
\end{equation}
where $f_j$ is the flux of source $j$, the corresponding entry of
$\boldsymbol f$, and $\alpha$ sets the strength of the penalty. We set $w_j$ to zero for the protected
\lassocut{} targets and to unity for other sources; the backgrounds are
unpenalized. We also set
\begin{equation}
 \alpha=\sqrt{2\ln[\max(p,2)]},
 \label{eq:lasso-alpha}
\end{equation}
where $p$ counts, for each tile, the unprotected sources whose
model profiles overlap at least one pixel used in the fit ($G_{jj}>0$),
before sparse selection.

The fit has two steps. The first step minimizes $\chi^2/2+\mathcal P_1$
and selects the sources whose fluxes remain nonzero. It constrains all
fluxes to be nonnegative because a blended pair could otherwise split into a
positive and a negative component. The second step refits the selected
sources without the penalty to remove the shrinkage of their fluxes toward
zero. This refit also allows negative protected fluxes, so that faint
targets are not clipped at zero. Its covariance assumes a fixed set of selected sources, ignoring
the possibility that noise changes which fluxes are set to zero.

For the \lassocut{} targets of the comparison sample, the full-depth
sparse fit shows no significant change in accuracy relative to the
baseline. On the \lassofullncommon{} of these targets in its paired sample
(Appendix~\ref{app:sample-audit}), the sparse fit minus the baseline gives $\Delta\signmad=\lassofulldnmad$ (95\% interval
\lassofulldnmadci{}) and $\Delta\etaout=\lassofulldcat$ percentage points
(95\% interval \lassofulldcatci{}); \lassofulldiscorda{} targets are
catastrophic only in the sparse fit, against \lassofulldiscordb{} only in
the baseline. Compared with the same baseline, the sparse fit at \zcut{}
also leaves \signmad{} unchanged ($\Delta\signmad=\lassodnmad$, \lassodnmadci{}) but adds
catastrophic outliers ($\Delta\etaout=\lassodcat$ percentage points,
\lassodcatci{}; \lassodiscorda{} against \lassodiscordb{}).

The full-depth sparse fit gives calibrated redshifts to a slightly
larger fraction of the protected targets after the median-channel
\snpresel{} selection, with a difference from the baseline of
\lassofulldretention{} percentage points (\lassodretention{} at \zcut{}). However, this median uses only the channels available to the
redshift fit. If low-S/N channels are excluded after their fluxes are set
to zero, the median can rise without improving any measurement.
We therefore also compare the samples that satisfy \crit{}
(Equation~\ref{eq:criterion}), which counts channels instead of taking a
median. Excluding a channel can only lower this count, and channels whose
fluxes are set to zero therefore cannot help a source satisfy \crit{}.
With this criterion, the full-depth sparse fit selects roughly as many
protected targets as the baseline
(bottom left panel of Figure~\ref{fig:retention}).

However, the additional photometric redshifts obtained at full catalog
depth are often inaccurate. Within the comparison sample, which also
contains penalized sources fainter than \lassobound{}, \lassogainn{} sources
enter the valid sample (Appendix~\ref{app:sample-audit}) of the full-depth
sparse fit but not that of the baseline. These sources have $\signmad\simeq\lassogainnmad$ and
\lassogaincat{} catastrophic outliers and account for most of the outliers
in the bottom right panel of Figure~\ref{fig:zzarms}. Over the whole paired sample of
\lassobothn{} sources, which also includes the penalized sources, the catastrophic fractions are
\lassobothcatlasso{} for the full-depth sparse fit and
\lassobothcateig{} for the baseline. Most of this excess comes from the
\lassopenn{} penalized sources in the paired sample, for which \signmad{}
rises from \lassopennmadeig{} to \lassopennmadlasso{} and the catastrophic
fraction from \lassopencateig{} to \lassopencatlasso{}, whereas the protected
targets alone differ by only \lassofulldcat{} percentage points. The same
degradation appears in the faintest magnitude bin of
Figure~\ref{fig:retention}, which contains these penalized sources (bottom
right panels).

These results hold for the current implementation, whose covariance
estimate omits the uncertainty in which penalized sources are selected.
Additional tests should therefore propagate that uncertainty into the
redshift fits of the penalized sources and assess the resulting accuracy.
Until then, where a reliable external catalog defines
the depth cut, we recommend truncation over the sparse configuration
tested here.

\section*{Data Availability}

Every input is a public product cited in Section~\ref{sec:data}.
\tractorjax{} is open source under the terms of \thetractor{}, of which it
is a derivative, at \url{https://github.com/hbahk/tractor-jax}; the release
behind the measurements in this paper is archived at Zenodo
\citep[v0.3.0;][]{bahk2026tractorjax}. The
\spherex{} layer of Section~\ref{sec:model} is at
\url{https://github.com/hbahk/tractorjax-spherex}, release v0.3.1, which
requires that \tractorjax{} release. The derived products and analysis
settings used in this work are available upon request.

\bibliographystyle{aasjournalv7.1}
\bibliography{references}

\end{document}